\documentclass[preprint,12pt]{elsarticle}

\usepackage{amssymb}
\usepackage{amsmath}
\usepackage{xcolor}

\journal{}

\begin{document}

\begin{frontmatter}

\title{Spatio-temporal pulse propagation during highly-resolved onset of Rayleigh-Taylor and Kelvin-Helmholtz Rayleigh-Taylor instabilities}

            \author[1]{Bhavna Joshi}

		\author[1]{Aditi Sengupta\corref{corr-auth}}
		\ead{aditi@iitism.ac.in}

       	\author[2]{Yassin Ajanif}
        \author[2]{Lucas Lestandi}

		\affiliation[1]{organization={Department of Mechanical Engineering, Indian Institute of Technology (Indian School of Mines)},
			city={Dhanbad},
			postcode={826004}, 
			state={Jharkhand},
			country={India}}
   
		\affiliation[2]{organization={École Centrale Nantes, Nantes Université, CNRS, GeM, UMR 6183, F-44000 Nantes, France},
            }

\begin{abstract}
The present study explores the onset of the Rayleigh-Taylor instability (RTI) and Kelvin-Helmholtz Rayleigh-Taylor instability (KHRTI) with highly-resolved direct numerical simulations of two setups which consider air at different temperatures (or densities) and/or velocities in two halves of three-dimensional cuboidal domains. The compressible Navier-Stokes equations are solved using a novel parallel algorithm which does not involve overlapping points at sub-domain boundaries. The pressure disturbance field is compared during onset of RTI and KHRTI and corresponding convection- and advection-dominated mechanisms are highlighted by instantaneous features, spectra, and proper orthogonal decomposition. The relative contributions of pressure, kinetic energy and rotational energy to the overall energy budget is explored for both instabilities, revealing acoustic trigger to be the incipient mechanism for both RTI and KHRTI. The nonlinear, spatio-temporal nature of the instability is further explored by application of a transport equation for  enstrophy of compressible flows. This provides insights into the similarities and differences between the onset mechanisms of RTI and KHRTI, serving as a benchmark data set for shear and buoyancy-driven instabilities across diverse applications in geophysics, nuclear energy and atmospheric fluid dynamics. 
\end{abstract}


\begin{highlights}
\item Comparison of onset mechanisms for buoyancy-driven Rayleigh-Taylor instability (RTI) and shear-buoyancy-driven Kelvin-Helmholtz Rayleigh-Taylor instability (KHRTI) via acoustic pulse propagation and vorticity dynamics.
\item Convection-dominated RTI and advection-dominated KHRTI evoke interface normal and radial propagation of pulses, respectively, in the plane of shear addition.
\item Anomalous modes are responsible for transient effects during growth of RTI and KHRTI while regular modes are observed in pairs while performing proper orthogonal decomposition.
\item Acoustics trigger the RTI/KHRTI, followed by creation of baroclinic torque.
\item Viscous stresses are dominant for the growth of RTI and KHRTI during onset, with baroclinicity and compressibility gaining prominence.
\end{highlights}


\end{frontmatter}



\section{Introduction}
\label{sec1}

Hydrodynamic instabilities and associated disturbance growth occur across a wide range of scales \cite{zhou2021rayleigh}, from controlled laboratory experiments \cite{read1984experimental, browand1973lab} to large-scale phenomena in astrophysics, geophysics, and cosmology \cite{cabot2006reynolds, Rayleigh2}. Even minor deviations from equilibrium can trigger such growth, which is often amplified by mechanisms such as buoyancy, shear forces, or convection (both natural and forced). Examples of hydrodynamic instabilities include buoyancy-driven Rayleigh-Taylor instability (RTI) and shear-induced Kelvin-Helmholtz instability (KHI), as well as their combined effect in Kelvin-Helmholtz-Rayleigh-Taylor instability (KHRTI). KHRTI frequently occurs in natural systems, such as oceans and atmospheres \cite{turner1979buoyancy, sengupta2023multi}, and in engineered environments like combustion chambers \cite{nagata2000effects} and inertial confinement fusion targets \cite{atzeni2004physics}. The interplay between shear and buoyancy in driving mixing layers has been explored in various contexts, including plasma physics \cite{shumlak1998mitigation} and classical fluid dynamics \cite{olson2011nonlinear}.

The RTI occurs when a denser or colder fluid rests atop a lighter or warmer one under the effect of gravity \cite{zhou2024hydrodynamic}. Initially static, the system destabilizes due to baroclinic torque, triggering convection and vortex formation \cite{sengupta2021role}. Rayleigh's early experiments \cite{Rayleigh1} observed finger-like structures as warm dyed water descended into cold fresh water. Taylor \cite{taylor1950instability} later provided a theoretical framework for inviscid, incompressible fluids with sinusoidal interface perturbations, extended to viscous fluids \cite{chandrasekhar1961hydrodynamic}. Linear analyses predicted exponential growth at early stages, while later stages deviated from linear theory \cite{sharp1984overview}. Youngs \cite{youngs1989modelling} predicted quadratic mixing layer growth, confirmed by Read's experiments \cite{read1984experimental}. Research on gravitationally accelerated RTI \cite{andrews1990simple} and forced versus unforced scenarios \cite{roberts2016effects} has deepened understanding of interior mixing dynamics. Studies have linked RTI growth to density or thermal stratification \cite{sharp1984overview, sengupta2023role}, Reynolds number \cite{cabot2006reynolds}, and fluid miscibility \cite{roberts2016effects}. Sengupta {\it et al.} \cite{sengupta2022three} used a high-resolution three-dimensional (3D) direct numerical simulation (DNS) with 4.19 billion grid points to capture RTI onset and dynamics. Comprehensive reviews of RTI experiments, simulations, and models are available in \cite{zhou2017a_rayleigh}. The source of RTI is often attributed to the generation of baroclinic torque \cite{lawrie2010rayleigh}, but a recent study \cite{sengupta2022three} attributes the genesis of RTI to the acoustic pulse propagation. Here, we will examine and compare both disturbance fields and answer the question - what comes first: pressure pulse or baroclinic vorticity?

The KHI occurs when two stratified fluids flow tangentially at an interface with differing velocities ($\Delta U$) \cite{zhou2024hydrodynamic}. Early experiments \cite{liepmann1947investigations} revealed that maximum kinetic energy production, diffusion, and dissipation occurred at the mixing layer's center. Mixing layer growth in KHI is driven by vortex pairing, where neighboring vortices merge to form larger ones, as visualized in dye-based water tunnel experiments \cite{winant1974vortex}. Studies show increased entrainment of fluid into vortical structures as the mixing layer transitions from laminar to turbulent \cite{koochesfahani1986mixing}, with enhanced mixing due to three-dimensional effects post-transition. Research on shear-driven mixing between fluids with density gradients found fluid entrainment within vortices but limited influence of density on mixing layer growth \cite{brown1974density}. A detailed review of turbulent mixing and transition criteria for RTI, KHI, and Richtmeyer-Meshkov instability is available in \cite{zhou2019turbulent}.

Experiments with stably stratified free shear flows in water channels \cite{browand1973lab} showed that stratification suppresses vortex pairing, reducing turbulence production. In KHRTI scenarios with varying velocity differences ($\Delta U$) and constant density gradients, a transition from homogeneous shear layer growth (for negligible density gradients relative to $\Delta U$) to quadratic mixing layer growth was observed \cite{lawrence1991stability}. Initially, shear-driven KHI dominates, but buoyancy-driven RTI takes over as mixing progresses. Adding shear has minimal effect on growth once buoyancy dominates \cite{finn2014experimental, akula2017dynamics, sengupta2023effects}. While RTI's baroclinic term and KHI's shear are known triggers, the precise onset mechanisms for KHRTI remains under investigation. Numerical studies \cite{sengupta2023effects, sengupta2022thermally} reveal that the imposed shear and thermal gradient determine whether the instability begins via RTI (characterized by billowing from the inflow plane) or KHI (marked by KH eddies). A transition between these mechanisms exhibits distinct turbulence scalings: -11/5 for buoyancy-dominated KHRTI and -5/3 for shear-dominated KHRTI. Recent work \cite{joshi2024highly} analyzed the onset of KHRTI with an enstrophy transport equation, highlighting the dominance of viscous stress terms using a high-resolution DNS with 480 million grid points. The present study compares the onset of RTI and KHRTI via pressure perturbation analysis, drawing insights into acoustic pulse evolution and growth in the presence and absence of shear. Furthermore, a proper orthogonal decomposition (POD) is conducted, which has not been reported for the onset stages of RTI and KHRTI before. 

The current study utilizes highly-resolved 3D DNS of RTI and KHRTI to examine instability onset through detailed analysis of pressure perturbations, spectra, POD modes, and the compressible enstrophy budget. By solving the compressible Navier-Stokes equations (NSE), we investigate the role of traveling pressure perturbations in triggering RTI and KHRTI, emphasizing the similarities and differences in the underlying mechanisms. The correlation of POD modes with those from internal flows and bluff bodies \cite{noack2003hierarchy} provides insights into spatiotemporal dynamics of buoyancy-driven instabilities and the influence of shear on the pressure field. The paper is structured as follows: Section \ref{sec2} outlines the problem formulation and numerical methods. Results are discussed in section \ref{sec3}, including analysis of pressure pulses via instantaneous contours in two perpendicular planes and corresponding spectra. The POD analysis details amplitude functions and spatial eigenmodes during onset of the RTI and KHRTI. The vorticity dynamics is compared and contrasted with the pressure perturbation, answering once and for all what is the actual trigger for RTI/KHRTI - acoustics or baroclinic torque. The CETE highlights the common dominant mechanisms for RTI and KHRTI and how sub-dominant budget terms vary during disturbance growth. The paper concludes with a summary and conclusions in Section \ref{sec4}. 

\section{Problem formulation for the RTI and KHRTI}
\label{sec2}

\begin{figure*}[!ht]
\centering
\includegraphics[width=0.9\textwidth]{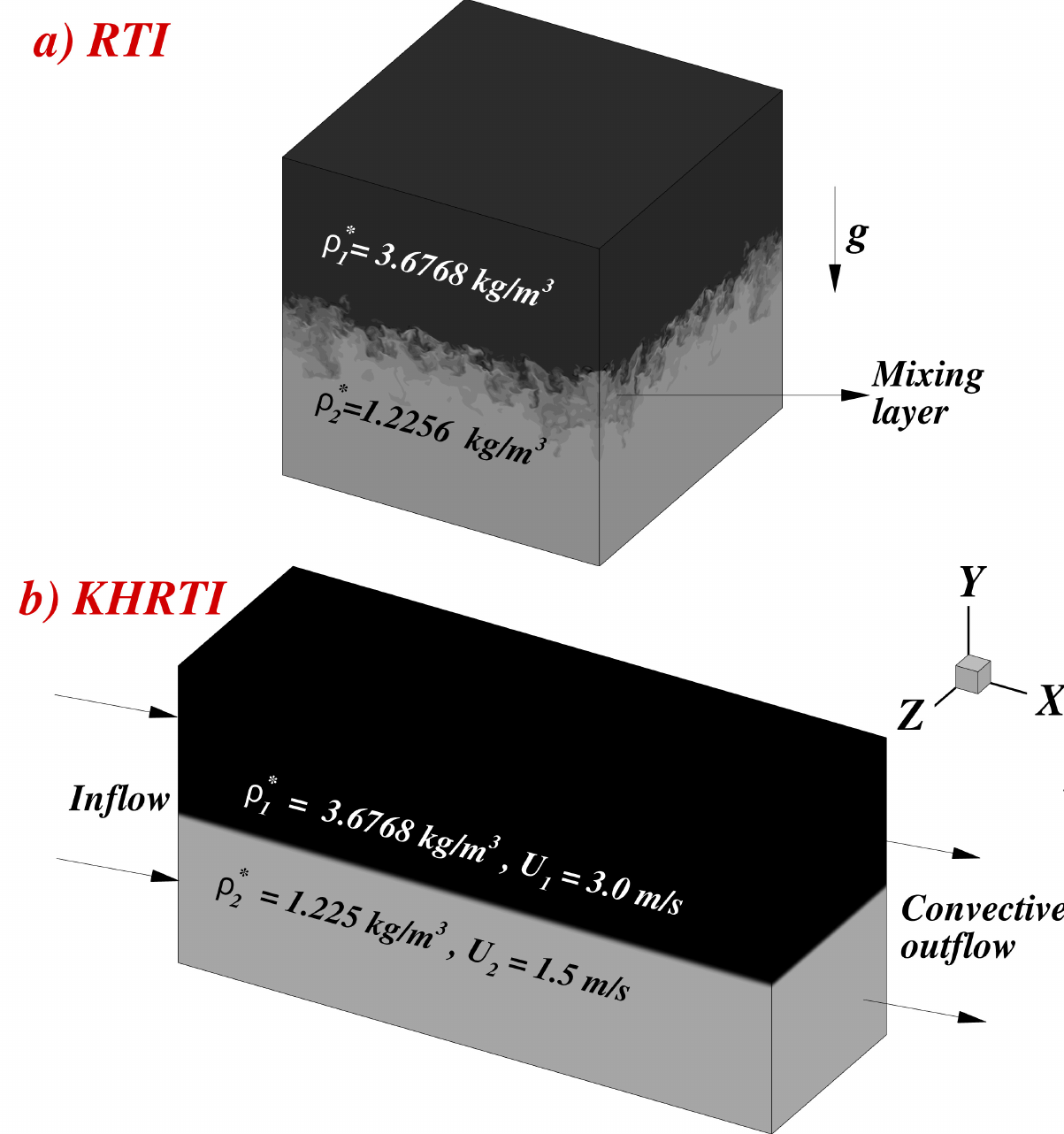}
\caption{Schematic of the computational domain for (a) RTI and (b) KHRTI, with labelled boundary conditions.}
\label{fig1}
\end{figure*}

The RTI simulation is performed within a three-dimensional (3D) cubic domain, each side measuring $L = 0.15$m, as depicted in Fig. \ref{fig1}(a). The computational domain is uniformly discretized with 1280 grid points along each direction of the horizontal ($x$,$z$)-plane and 2560 grid points along the vertical ($y$)-direction, which is perpendicular to the initial interface. Initially, the domain is divided into two compartments separated by an insulated planar partition. The lower compartment contains air at a constant temperature, $T^*_l = T^*_s = 300 K$, while the upper compartment contains colder air at $T^*_u = 100 K$. The corresponding air densities are determined using the ideal gas law. For both the RTI and KHRTI, air density at 300K is taken as the density scale, $\rho_s = 1.2256 kg/m^3$. The configuration and parameters closely follow the experimental setup described by Read \cite{read1984experimental}. All walls of the domain are adiabatically insulated, ensuring the system remains thermodynamically isolated. No-slip boundary conditions are imposed on all boundaries. At $t = 0$, the separating partition is removed, initiating the RTI. The instability is further influenced by acoustic disturbances, which are described in detail in the subsequent section.

The numerical simulation for the KHRTI is carried out within a 3D cuboidal domain of dimensions $(L \times H \times W) = (0.3 m \times 0.15 m \times 0.1 m)$, as illustrated in Fig. \ref{fig1}(b). The computational grid comprises 1501 points along the length ($L$), 801 points along the height ($H$), and 401 points along the width ($W$). This domain size and shape are inspired by the experimental setup described in \cite{akula2017dynamics}, where a larger cuboid ($3m \times 1.8m \times 0.6m$) was used to visualize the KHRTI. The flow configuration involves an upper fluid layer moving with a streamwise velocity of $U_1 = 3 m/s$ and a lower layer moving at $U_2 = 1.5 m/s$. At the left boundary, an inflow condition is imposed, while a convective outflow condition is applied at the right boundary. Non-reflective boundary conditions based on Riemann invariants \cite{sengupta2013high} are utilized at the inflow to account for upstream-propagating acoustic components while solving the compressible NSE. The inflow pressure and density are computed using the relations, $p = \rho a^2/\gamma$ and $log(p/\rho^\gamma) = s$, where $a$ is the speed of sound, $\gamma$ is the adiabatic index, and $s$ is the entropy. At $t=0$, a finite-thickness partition separating the two fluid streams is impulsively removed in the horizontal direction, initiating the mixing process. This removal serves as the initial perturbation, triggering instability across a range of spatial and temporal scales driven by destabilizing potential energy. Notably, no explicit perturbations are applied at the interface, mimicking unforced experimental conditions reported in studies of hydrodynamic instabilities \cite{read1984experimental, akula2017dynamics, roberts2016effects}. The observed instability results from the interplay of two opposing mechanisms: a statically unstable configuration with heavier fluid overlying lighter fluid, and shear induced in the streamwise direction.

The numerical simulations of the RTI and KHRTI are performed by solving the unsteady 3D compressible NSE \cite{hoffmann2000computational} given as:

\begin{equation}
\frac{\partial \hat{Q}}{\partial t^*}+\frac{\partial \hat{E_c}}{\partial x^*}+\frac{\partial \hat{F_c}}{\partial y^*}+\frac{\partial \hat{G_c}}{\partial z^*}=\frac{\partial \hat{E_v}}{\partial x^*}+\frac{\partial \hat{F_v}}{\partial y^*}+\frac{\partial \hat{G_v}}{\partial z^*} + \hat{S},
\label{ge1}
\end{equation}

\noindent for the conserved variables given as
\begin{equation}
\hat{Q} = [ \rho^*; \; \rho^* u^*; \; \rho^* v^*; \; \rho^* w^*; \; \rho^* e_t^* ]^T.
\label{ge2}
\end{equation}
	
\noindent The convective flux variables $\hat{E}_c$, $\hat{F}_c$ and $\hat{G}_c$ are given as

\begin{equation}
\hat{E}_c = [ \rho^* u^*; \; \rho^* u^{*2} + p^*; \; \rho^* u^* v^* ; \; \rho^* u^* w^*; \; (\rho^* e_t^* + p^*) u^*]^T,
\label{ge3}
\end{equation}
 
\begin{equation}
\hat{F}_c = [ \rho^* v^*; \; \rho^* u^* v^*; \; \rho^* v^{*2} + p^*; \; \rho^* v^* w^*; \; 
(\rho^* e_t^* + p^*) v^* ]^T,
\label{ge4}
\end{equation}

\begin{equation}
\hat{G}_c = [ \rho^* w^*; \; \rho^* u^* w^*; \; \rho^* v^* w^*; \; \rho^* w^{*2} + p^*; \; (\rho^* e_t^* + p^*) w^* ]^T,
\label{ge5}
\end{equation}
	
\noindent and the viscous flux vectors $\hat{E}_v$, $\hat{F}_v$ and $\hat{G}_v$ are given as:

\begin{equation}
\hat{E}_v = [ 0; \; \tau^*_{xx}; \; \tau^*_{xy}; \; \tau^*_{xz}; \; u^* \tau^*_{xx} + v^* \tau^*_{xy} + w^* \tau^*_{xz} - q^*_x ]^T,
\label{ge6}
\end{equation}

\begin{equation}
\hat{F}_v = [ 0; \; \tau^*_{yx}; \; \tau^*_{yy}; \; \tau^*_{yz}; \; u^* \tau^*_{yx} + v^* \tau^*_{yy} + w^* \tau^*_{yz} - q^*_y ]^T,
\label{ge7}
\end{equation}

\begin{equation}
\hat{G}_v = [ 0; \; \tau^*_{zx}; \; \tau^*_{zy}; \; \tau^*_{zz}; \; u^* \tau^*_{zx} + v^* \tau^*_{zy} + w^* \tau^*_{zz} - q^*_z ]^T.
\label{ge8}
\end{equation}

\noindent The source term in Eq. \eqref{ge1} due to buoyancy, $\hat{S}$ is given by

\begin{equation}
\hat{S} = [ 0; \; 0; \; -\rho^*g; \; 0; \; -\rho^* v^* g ]^T.
\label{source}
\end{equation}
	
Flow variables $\rho^*$, $p^*$, $u^*$, $v^*$, $w^*$, $T^*$ and $e_t^*$ represent dimensional density, pressure, Cartesian components of velocity, the absolute temperature and specific internal energy, respectively. The stress tensor $\tau^*_{ij}$, for $i,j = 1$ to 3, is related to the rate of strains as,

\begin{equation}
\tau^*_{ij} = \tau^*_{ji} =  \mu^* \biggl( \frac{\partial v_{j}^*}{\partial x_{i}^*} +  \frac{\partial v_{i}^*}{\partial x_{j}^*} \biggr );
		\; \tau^*_{ij} = \biggl[ \mu^* \biggl (\frac{\partial v_{j}^*}{\partial x_{i}^*} +  \frac{\partial v_{i}^*}{\partial x_{j}^*} \biggr ) + \lambda^* \frac{\partial v_{i}^*}{\partial x_{i}^*}   \biggr ] \; {\rm for} \; i = j;
\label{stress}
\end{equation}
	
\noindent The specific heat capacity ($C_v$) and thermal conductivity ($\kappa$) are constants in this formulation. The heat conduction terms $q^*_i$ are given as: $q^*_{i} = - \kappa \frac{\partial T^*}{\partial x^*_{i}}$. The bulk viscosity $\mu_b = \lambda^* + 2\mu^*/3$ is obtained by performing a regression analysis of the experimental data in \cite{ash1991second}. Sutherland's law is used for evaluating the dynamic viscosity ($\mu^*$) as a function of temperature ($T^*$). Additionally, the ideal gas relation, $p^*=\rho^* R^*T^*$ provides the constitutive equation. This in turn defines specific energy, $e_t^*$ with summation convention of Einstein used in the relation as
	
\begin{equation}
e_t^* =\frac{p^*}{\rho^*(\gamma-1)} + \frac{v_{i}^*v_{i}^*}{2}.
\label{ge13}
\end{equation}
	
\noindent The independent and dependent variables are nondimensionalized \cite{hoffmann2000computational} as follows,
	
\begin{equation*}
x_i = \frac{x^*_{i}}{L}, \; v_i = \frac{v^*_{i}}{U_s}, \;t = \frac{t^*U_s}{L}, \; \rho = \frac{\rho^*}{\rho_s}, \; p = \frac{p^*}{p_s},\; e_t = \frac{e^*_t}{U^2_s}, \; T = \frac{T^*}{T_s}, \; \mu = \frac{\mu^*}{\mu_s}
\end{equation*}
 
\noindent where $L$, $U_s$, $T_s$ and $\rho_s$ are the length, velocity, temperature, and density scales and $p_s = \rho_sR^*T_s$, is the characteristic pressure obtained with the reference temperature and density for hot air. For KHRTI, the velocity scale, $U_s = \sqrt{gL}$, and length scale, $L$ are chosen to be $1.5m/s$ and $0.15m$, respectively. The corresponding time scale is $0.1s$. For RTI, we define velocity and time scales as $U_s = 12.131$m/s and 0.1237s, respectively. For both RTI and KHRTI, we have used $\Delta T= 200K$, and viscosity scale, $\mu_s$ as viscosity of air at $T_s$. Flow similitude is expressed in terms of characteristic parameters: Reynolds number ($Re$), Prandtl number ($Pr$), Froude number ($Fr$), and Gay-Lussac number ($Ga$) as,

\begin{equation}
Re = \frac{\rho_s U_s L}{\mu_s}, \; Pr = \frac{\mu C_p}{\kappa}, \; Fr = \sqrt{\frac{U^2_s}{gL}}, \; Ga = \frac{\Delta T}{T_s}.
\label{ge16}
\end{equation}

\noindent Here, $Ga$ relates temperature difference from the reference temperature ($T_s = 300 K$) and is equal to $0.667$ for both the RTI and KHRTI. In the present computations, we have used $Pr = 0.712$.  With a perfect gas as the working fluid in both the compartments, the corresponding Atwood number [$At = (\rho_1 - \rho_2)/(\rho_1 + \rho_2)$] is $At = 0.5$. The reference Reynolds number and the Froude number for the KHRTI are $Re = 15000$, $Fr = 1.2365$, respectively. The reference $Re$ for RTI is 12080.6 and $Fr = 1.6899$. The non-dimensional governing equations are thus obtained as,
 
\begin{equation}
\frac{\partial Q}{\partial t}+\frac{\partial E_c}{\partial x}+\frac{\partial F_c}{\partial y}+\frac{\partial G_c}{\partial z}=\frac{\partial E_v}{\partial x}+\frac{\partial F_v}{\partial y}+\frac{\partial G_v}{\partial z} + \hat{S}, 
\label{ge17}
\end{equation}

\noindent where the normal stress components are (for $i = j$):

\begin{equation}
\tau^*_{ij} = \frac{1}{Re} \biggl[ \mu^* \biggl (\frac{\partial v_{j}^*}{\partial x_{i}^*} +  \frac{\partial v_{i}^*}{\partial x_{j}^*} \biggr ) + \biggl (\frac{\mu_b}{\mu_s} - \frac{2\mu}{3} \biggr ) \frac{\partial v_{i}^*}{\partial x_{i}^*}   \biggr ]
\end{equation}

\noindent and the shear stress terms are given by 
	
\begin{equation}
\tau^*_{ij} = \tau^*_{ji} =  \frac{\mu}{Re} \biggl( \frac{\partial v_{i}}{\partial x_{j}} +  \frac{\partial v_{j}}{\partial x_{i}} \biggr ).    
\end{equation}
	
\noindent The heat conduction terms are given as follows

\begin{equation}
q_i = - \frac{\mu}{(\gamma-1)PrReM^2}\frac{\partial T}{\partial x_i}.
\label{ge20}
\end{equation}
	
The equation of state can be written in its nondimensional form as $p = \rho RT$, where $R = R^* T_s / U_s^2$. Past simulations of RTI \cite{sengupta2016roles, sengupta2021role, sengupta2016non} have shown the necessity to include the bulk viscosity to capture the early-time behavior of RTI. Here, we use acoustic dispersion and attenuation measurements \cite{ash1991second} which provide values of $\mu_b$ for air as a function of temperature. A regression analysis of the experimental data provides a relationship between $\mu_b$ and $T^*$ (in $K$) as, $\mu_b = \frac{1}{10^4}[3.381\times T^* - 7.383 ]$.

\subsection{Numerical methodology} 
\label{subsec:num_meth}

For the KHRTI simulations, a non-uniform compact scheme is applied in the physical plane, following the approach described in \cite{sengupta2017hybrid}. The computational grid uses uniform spacing in the $x$- and $z$-directions, while non-uniform spacing is adopted in the $y$-direction. This non-uniformity provides refined resolution near the top and bottom walls as well as at the interface between the two fluid streams at 
$t=0$. The primary focus of this study is to investigate the early-stage dynamics of the KHRTI by monitoring the evolution of pressure perturbations. The refined mesh near the walls and the interface effectively resolves the propagation of pressure pulses during the initial phase of the instability. In contrast, the RTI computations employ a uniformly spaced grid in all directions.

The KHRTI simulation is executed with 1200 cores using a mesh consisting of 480 million grid points, while the RTI simulation employs 19,200 cores to handle a much more refined grid with 4.19 billion points. These simulations, performed at the peta-scale level, utilize advanced numerical schemes to ensure high accuracy and efficiency. For the sub-domain boundary closures in parallel programming, a dispersion-relation-preserving (DRP) compact scheme is used, as outlined in \cite{sundaram2022non}. The convective flux terms in the interior are discretized using different schemes tailored to the specific requirements of each instability. For KHRTI, a sixth-order non-uniform compact scheme (NUC6) \cite{sengupta2017hybrid} is employed, while RTI computations utilize an optimized upwind compact scheme (OUCS3) \cite{sengupta2013high}. Boundary closure in the KHRTI simulation is handled by an alternate bi-diagonal ADB3 scheme \cite{sengupta2016new}. For viscous flux calculations, a second-order central differencing scheme is used, adapted for non-uniform grids in KHRTI and uniform grids in RTI \cite{sagaut2023global}.

Time integration in both simulations is carried out using a four-stage, fourth-order Runge-Kutta method. The non-dimensional time steps are carefully selected to maintain numerical stability and accuracy: $\Delta t = 6.00 \times 10^{-8}$ for KHRTI and $\Delta t = 6.25 \times 10^{-7}$ for RTI. These values are chosen to preserve the numerical dispersion properties necessary for achieving neutral stability of the convective term while minimizing phase and dispersion errors. The selection of time steps is guided by the Courant-Friedrichs-Lewy (CFL) number, $(Nc = \frac{c\Delta t}{\Delta x})$ and diffusion number $(D_n = \frac{\nu \Delta t}{(\Delta x)^2})$. Here $c$ is the characteristic wave speed, $\nu$ is the kinematic viscosity, and $\Delta x$ represents the grid spacing. Use of non-uniform spacing in KHRTI yields a finer grid spacing in the wall-normal direction, thereby requiring a smaller time-step than RTI to preserve neutral stability. The lack of grid metric calculations in NUC6 ensures that the KHRTI simulation is faster despite the lower time-step when compared with an equivalent uniform spacing calculation.

\section{Results and Discussion}
\label{sec3}

In this section, we will explore the onset processes of RTI and KHRTI by tracking the contours of the disturbance pressure field in two perpendicular planes. Emphasis is placed on the disturbance pressure, as the early onset of RTI and KHRTI is triggered by acoustic pulses, as described in  \cite{sengupta2022three} and \cite{joshi2024highly}, respectively, and in other references cited therein. The corresponding spectra are compared here for the early onset and propagation stages of RTI and KHRTI in the following subsection. The differences in the onset mechanisms of these spatio-temporal growths are furthermore substantiated through a proper orthogonal decomposition of the disturbance pressure. Finally, role of rotationality is examined by tracking the disturbance vorticity in ($x$, $y$)- and ($y$, $z$)-planes. We compare translational kinetic energy budget with rotational compressible enstrophy budget is explored.

\subsection{Onset of RTI versus KHRTI: Analysis of pressure perturbation field}

First, we show the non-dimensional disturbance pressure field in two perpendicular planes: ($x$, $y$) and ($y$, $z$)-planes for RTI and KHRTI, computed in domains shown in Figs. \ref{fig1}(a) and (b), respectively. The disturbance field is obtained by subtracting the instantaneous pressure from the initial hydrostatic pressure. The non-dimensional pressure disturbance magnitudes associated with RTI and KHRTI are of the order of $10^{-6}$ during the early onset stage, which are considerably smaller compared to the unperturbed hydrostatic pressure. Thus, the use of highly accurate dispersion relation preserving methods enables us to capture such spatial scales in the flow.  

\begin{figure*}[!ht]
\centering
\includegraphics[width=.98\textwidth]{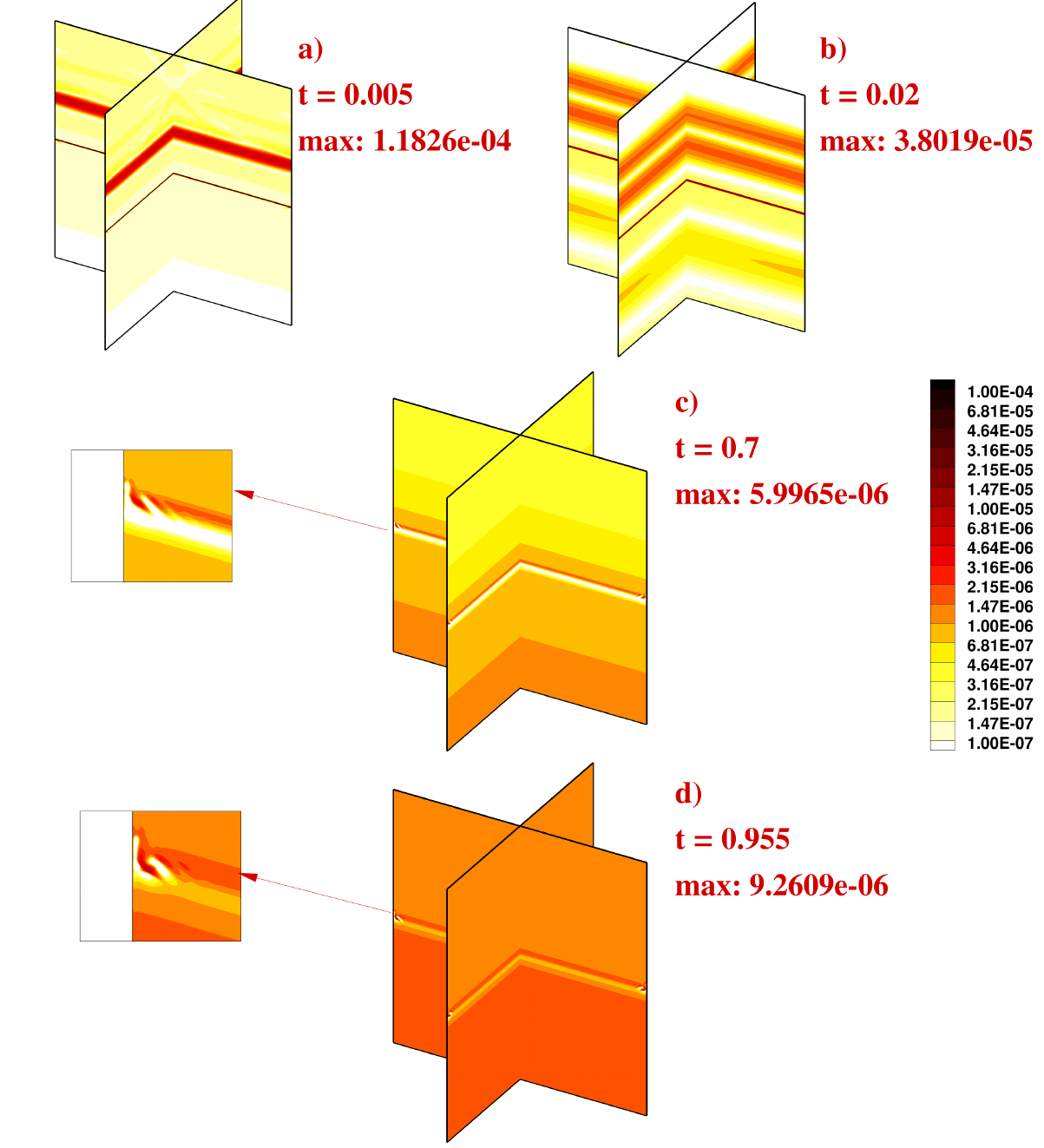}
\caption{Disturbance pressure contours of the onset of the RTI demonstrating primarily a 1D behavior.}
\label{fig2}
\end{figure*}

In Fig. \ref{fig2}, the evolution of disturbance pressure contours is demonstrated on ($x$, $y$) and ($y$, $z$)-planes for the onset of the RTI, at indicated time instants. For early time instants, such as $t = 0.005$ and $t = 0.02$, one observes interface normal pressure pulses in both ($x$, $y$) and ($y$, $z$)-planes. The travelling compression and rarefaction fronts originate from the interface and propagate on either direction towards the top and bottom walls, as has been observed in \cite{sengupta2022three} for early stages of RTI. Beyond $t = 0.7$ in Fig. \ref{fig2}(c), one observes the generation of viscous finger-like structures (as seen in the zoomed inset plots) from the juncture of the original interface with the two side-walls. This can be attributed to the generation of baroclinic vorticity triggered by the travelling acoustic pulses in earlier frames. For the RTI computed so far, the latest time frame is shown in Fig. \ref{fig2}(d), which shows the presence of much stronger viscous billows, an indicator of the early onset of RTI \cite{andrews1990simple}.

\begin{figure*}[!ht]
\centering
\includegraphics[width=.96\textwidth]{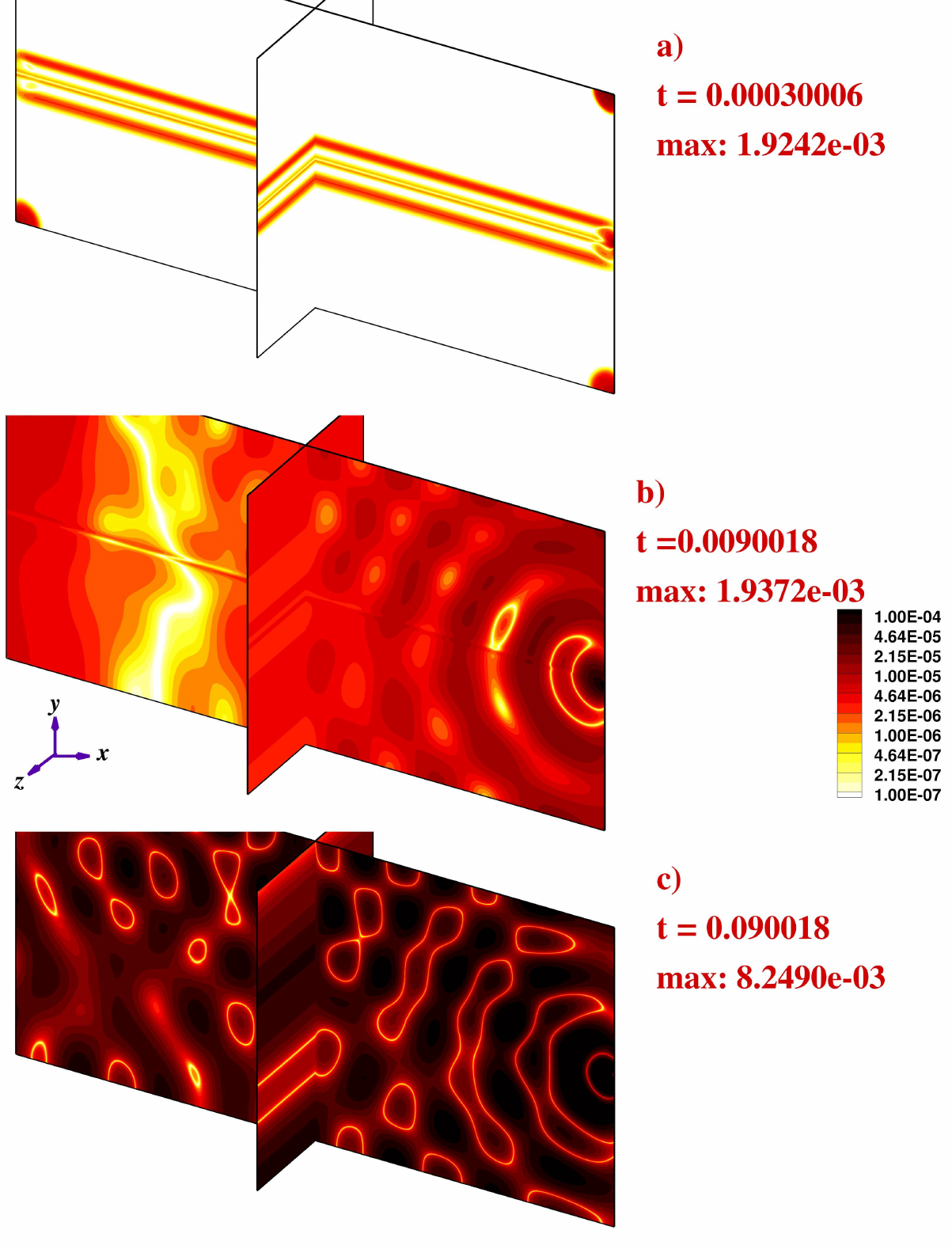}
\caption{Disturbance pressure contours for KHRTI during the onset showing radially traveling pulses.}
\label{fig3}
\end{figure*}

In Fig. \ref{fig3}, the evolution of disturbance pressure contours is demonstrated on ($x$, $y$) and ($y$, $z$)-planes for the onset of the KHRTI, at indicated time instants. Here, apart from the very early onset of KHRTI at $t = 0.0003$ in Fig. \ref{fig3}(a), the pressure pulses in the two perpendicular planes do not show a simple interface normal propagation as one noted for RTI in Fig. \ref{fig2}. Due to the imposed streamwise advection scale in KHRTI, beyond $t = 0.009$ in Fig. \ref{fig3}(b), a radially propagating pressure pulse is noted in the ($x$, $y$)-plane, originating from the outflow plane and convecting in the upstream direction. This follows the analytical expression of group velocity for KHRTI \cite{sengupta2012instabilities} during onset. At a subsequent time, $t = 0.09$ in Fig. \ref{fig3}(c), we observe radial upstream propagation of pressure pulses in the ($x$, $y$)-plane and interface-normal propagation of pressure pulses in the ($y$, $z$)-plane. Thus, for KHRTI, direction for pulse propagation is radial in plane containing the imposed shear gradient and normal in the perpendicular planes due to the imposed thermal gradient. 

We have examined the spatio-temporal pressure propagation in the physical planes for the RTI and KHRTI, distinguishing between the nature and direction of acosutic waves. Now, we further explore the spatio-temporal scales present in the flow for these two instabilities.  

\subsection{Spatio-temporal scale selection in RTI and KHRTI}

The RTI and KHRTI experiments conceived in the computational domains shown in Figs.\ref{fig1}(a) and (b), do not involve the imposition of any specific temporal or spatial scale. The discussion so far, however, demonstrates that depending upon the dominant mechanism (RT or KH), certain spatial scales in the flow are selectively amplified. This aspect is explored further in this section by computing the Fast Fourier transform (FFT) and recording the time evolution of the disturbance pressure signals. 

\begin{figure*}[!ht]
\centering
\includegraphics[width=.9\textwidth]{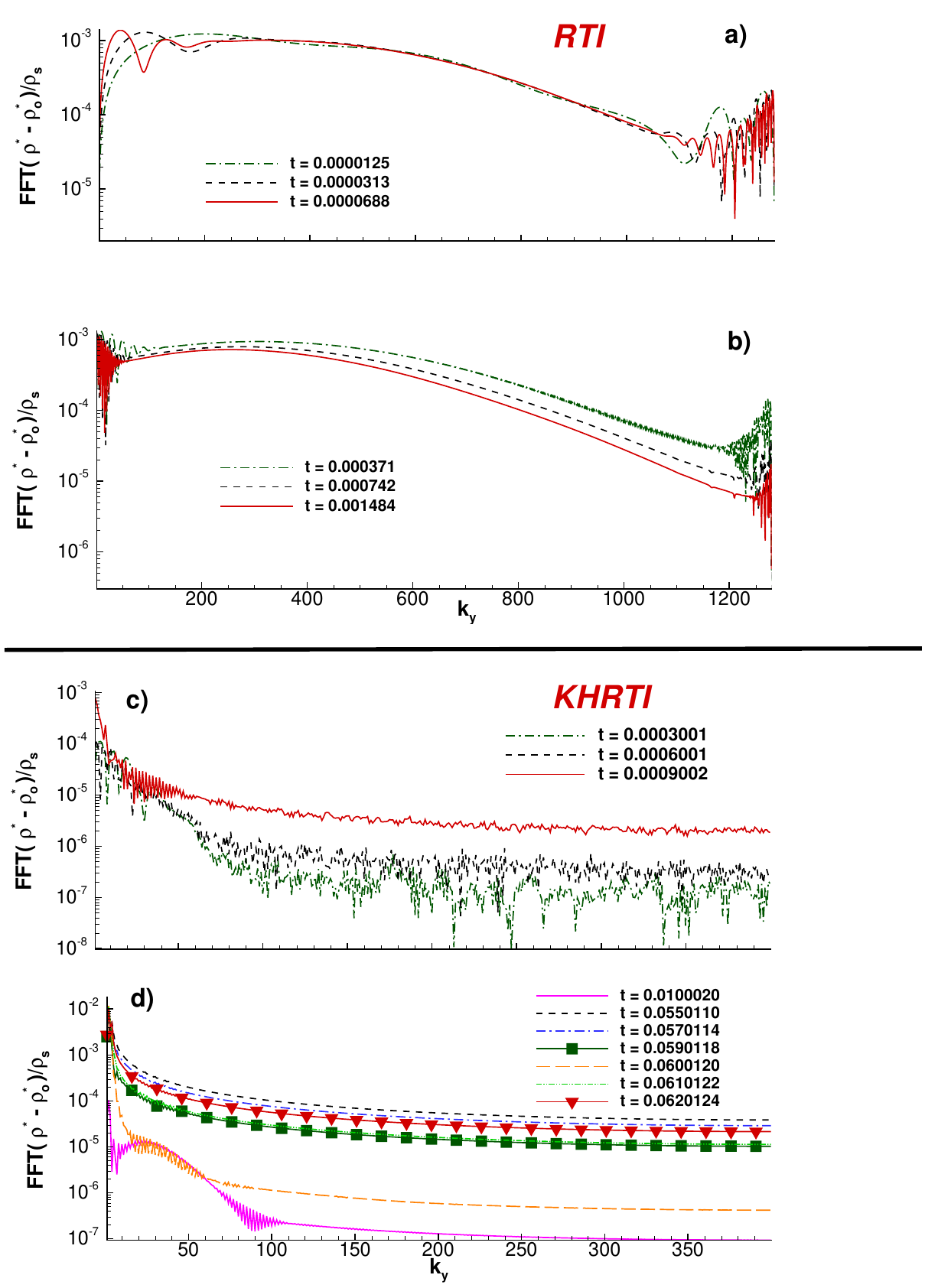}
\caption{Spectra of disturbance pressure at the indicated time instants during the (a) onset and at (b) later stages of RTI and (c) onset and at (d) later stages of KHRTI}
\label{fig4}
\end{figure*}

Figure \ref{fig4} shows the spectra of the disturbance pressure for the RTI at selected time instants. The signal variation is recorded with $y$, for a plane passing through $z = x = 0.5$. In Fig. \ref{fig4}(a), the FFT shows a typical nonmodal variation at low and moderate wavenumbers for $t = 0.0000125$, which is typical for transition to turbulence in external aerodynamics and turbomachinery \cite{sengupta2024separation}.  Interestingly, scale selection is observed to be near the highest non-dimensional wavenumber, $k_y$. In the next time instant, $t = 0.0000313$, in addition to the nonmodal hump, there is a superposition of a wave-packet near the origin of the wavenumber plane. These correspond to the low wavenumber components in the two halves of the computational domain, depicted in Fig. \ref{fig2}. One also notes a strong high wavenumber component at $t = 0.0000688$, which is representative of the collated high wavenumber component (formed due to interactions between low wavenumber pressure signals and their reflected counterparts) obtained in Fig. \ref{fig2}. 

\begin{figure*}[!ht]
\centering
\includegraphics[width=.82\textwidth]{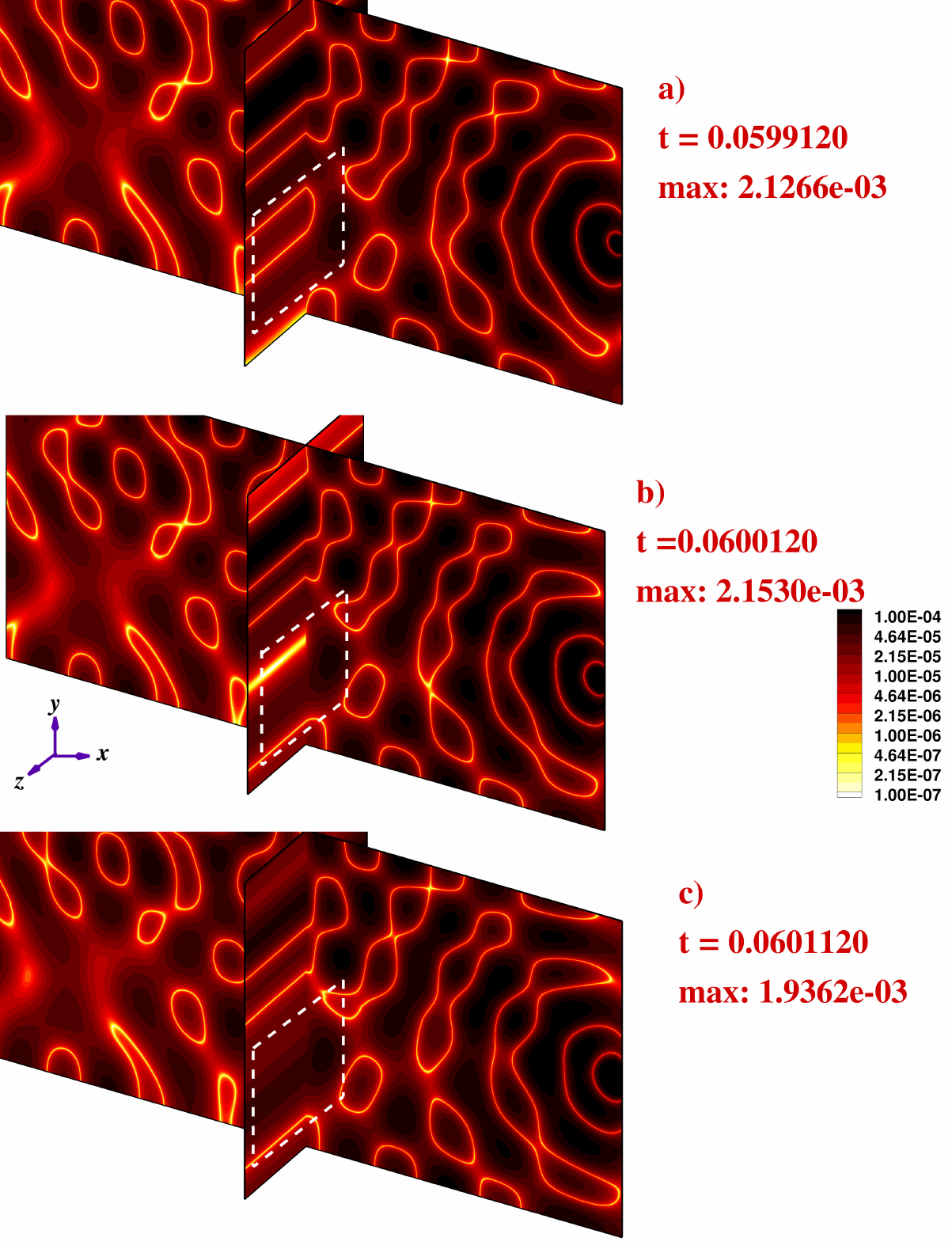}
\caption{Pressure disturbance contours explaining the anomalous behavior in spectra of disturbance pressure during later stages of KHRTI, shown in Fig.\ref{fig4}(d).}
\label{fig5}
\end{figure*}

Figure \ref{fig4}(b) shows the later stages of the RTI evolution with a clear demarcation between two parts of the spectrum: one near the origin with lower wavenumbers (including infrasounds and hydrodynamic pressure) and the other residing in the ultrasound range, at higher wavenumbers \cite{sengupta2022three}. The smooth local maxima representative of a nonmodal component is present at all times (near $k_y$ = 600) in addition to these two distinct spectrum components. This is considered to be the precursor to eventual transition to turbulence \cite{sengupta2020nonmodal}. For $t = 0.001484$, the modal signature of the spectrum near the origin of the wavenumber plane is replaced by a nonomodal low wavenumber wave-packet, as is observed in Fig. \ref{fig5}.  This suggests that the infrasound signal is transformed to ultrasound one, which transforms and attenuates with time. While the ultrasound becomes weaker at later times (not shown), the other component also decays. The infrasound component approaches significantly lower wavenumbers, those remaining alongside hydrodynamic pressure field. The nonmodal peak, also attenuates with time (from $t = 0.000371$ to 0.001484), transforming to a narrower peak. This is an ideal candidate for causing transition to turbulence for the RTI.

Figure \ref{fig4}(c) shows the spectra of the pressure disturbance for KHRTI at the selected time instants. The top frame shows the spectrum during the early onset of the KHRTI, where the interface-normal convection is the predominant motion for the pressure waves emanating from either side of the interface. The bottom frame shows the pressure spectra for later stages of the KHRTI where the streamwise advection of the KH mechanism sets in. Pressure disturbance is observed to be distributed over a larger range of wavenumbers compared to the RTI, which corroborates with the radial pressure pulses spanning over the whole domain in Fig. \ref{fig3}. The overall magnitude of the Fourier amplitude is higher for KHRTI than that of RTI, even for very early times. This is due to the higher disturbance levels triggered in KHRTI from the imposed shear. No distinct modal and nonmodal signatures are observed for KHRTI, right from the onset, in contrast to RTI. This chaotic nature of the pressure perturbation signal has a direct correlation with the interaction between convection-dominated 1D propagation of signal in RTI with the advection-dominated KH mechanism leading to an overall radial propagation of signal in KHRTI. 

During the later stages of the KHRTI shown in Fig. \ref{fig4}(d), a nonmodal peak is noted near $k_y$ = 40 at $t = 0.01$ which vanishes from $t = 0.055$ to $t = 0.059$. In fact, at the later times, the spectra's magnitude rises by three orders of magnitude with a pronounced peak near the origin of the wavenumber plane. Interestingly, at $t = 0.06$, the spectrum reverts back to the pattern and magnitude noted for the earlier time of $t = 0.01$. This anomalous reversion to the nonmodal peak's presence in the spectrum seems to suggest that the convection-dominated motion of the pressure signal (characteristic of RTI) takes over again. At subsequent stages, however, the spectra follow the typical behavior of the later stages of the KHRTI, wherein a radial pressure propagation leads to an overall increment in the Fourier amplitude of the pressure perturbation across all wavenumbers. To explain this anomalous behavior, we trace the contour plot of disturbance pressure over this time interval, in Fig. \ref{fig5}.

We plot the pressure perturbation contours in the ($x$,$y$) and ($y$,$z$)-planes for the indicated time instants (showing the anomalous behavior in the spectra of Fig. \ref{fig4}(d)) in Fig. \ref{fig5}. It is observed that radial pressure pulses propagate in the ($x$, $y$)-plane while compression and rarefaction fronts propagate in the interface-normal direction in the ($y$, $z$)-plane. At $t = 0.0599120$, in the dotted white box, there is a \lq mushroom-shaped' low wavenumber component in the bottom half of the domain of the order of $10^{-4}$. At $t = 0.0600120$, a rarefaction front from the interface to the bottom half of the domain is observed in the white dotted box. This is created by the destructive interference between the radial pulses in ($x$, $y$)-plane and pressure fronts convecting in ($y$, $z$)-plane. This leads to a sudden drop in magnitude of the pressure perturbation and for the spectrum to show RT-like behavior with a nonmodal peak due to convective propagation of pulses in interface-normal direction. At $t = 0.0601120$, this rarefaction front vanishes such that the spectrum of disturbance pressure reverts back to the typical one noted for the later stages of the KHRTI. This type of anomalous behavior is expected to occur periodically whenever there is an interaction between the radial pulses and the convective pressure fronts.

\begin{figure*}[!ht]
\centering
\includegraphics[width=0.83\textwidth]{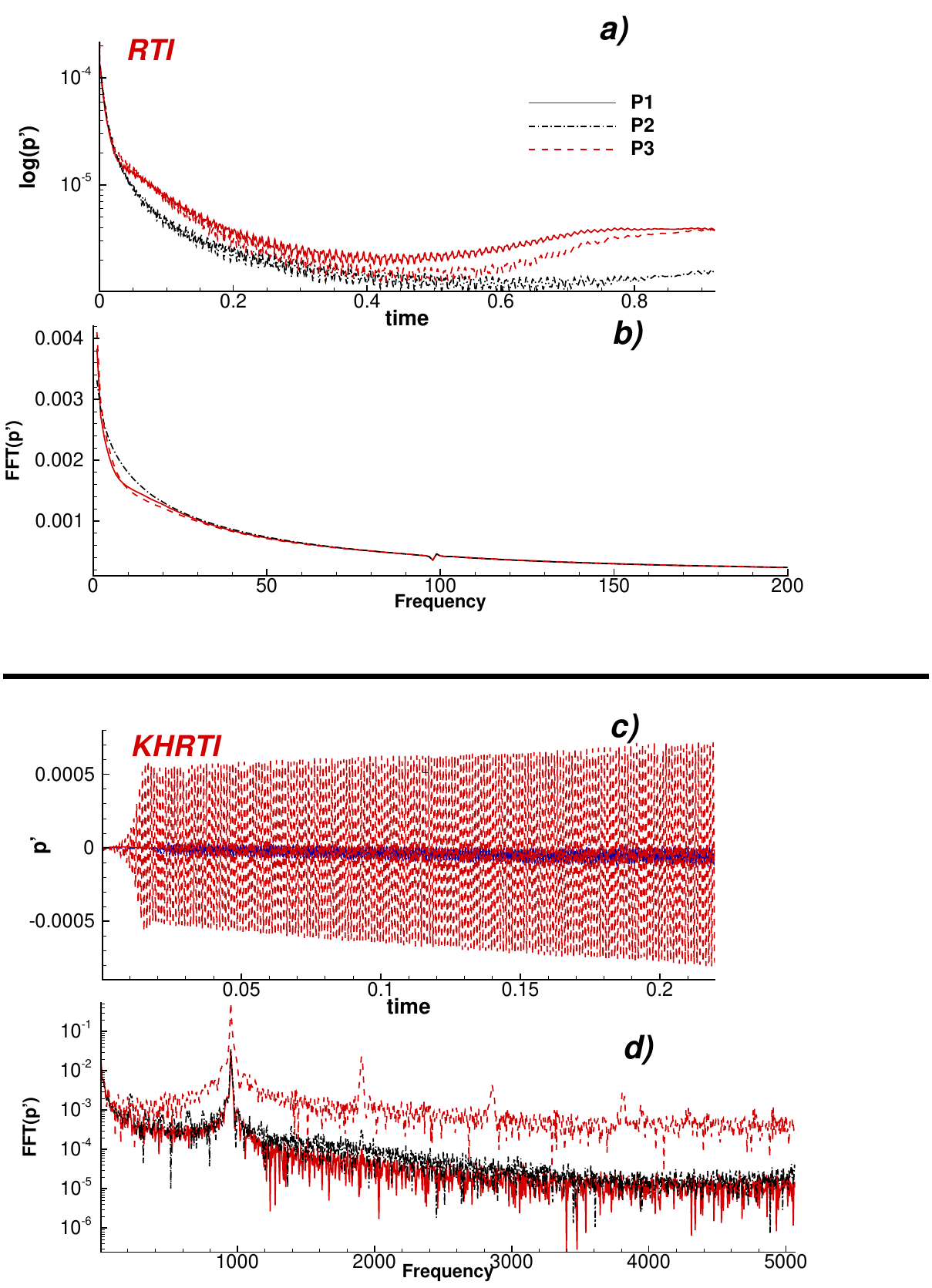}
\caption{Time-series of disturbance pressure probed at center (line b), left (line a), and right (line c) sides of original interface for RTI (top) and KHRTI (bottom). Corresponding spectra are also shown in the amplitude-frequency plane.}
\label{fig6}
\end{figure*}

The time-series of the pressure perturbation for RTI and KHRTI is recorded in Fig. \ref{fig6}(a) and \ref{fig6}(c) at three locations along the original interface of the $x = z = 0.5$-plane: (i) the left side (denoted by solid line p1), (ii) the centre (denoted by dash-dotted line p2), and (iii) the right side (denoted by dashed line p3). On a cursory glance, one notes that the order of magnitude of the disturbance pressure is lower at all locations for RTI compared to those for KHRTI. With time, $p'$ falls in case of RTI, especially once baroclinic vorticity generation begins from the side-walls. The acoustic pulses do not contribute significantly beyond providing the initial trigger for the instability. On the other hand, for KHRTI, the band of $p'$ grows with time, as the underlying mechanism responsible for the growth of the KHRTI is advection-dominated KH mechanism. Comparing the three probe locations, in RTI, the side-walls show a stronger disturbance pressure than the central location. This is expected as the junction points of the side-walls with the interface location are well-known to be trigger points for the RTI \cite{kucherenko2003experimental}. Symmetry exists for the side-wall locations till $t = 0.4$ beyond which creation of small-scale billowing disrupts the top-down symmetry \cite{sharp1984overview}. Beyond $t = 0.85$, the viscous fingering and billows formed lead to higher values of $p'$ (by an order of magnitude) than at the central location. It is expected that formation of spikes and bubbles at the centre during the later stages of RTI will show a pre-dominance of line \lq p2'. For KHRTI, the side-walls have similar order of magnitude of $p'$, which is orders of magnitude higher than the central location. The right outflow plane, in fact, has marginally higher values of $p'$ due to the upstream propagation of the radial pressure pulses. 

The corresponding spectra of the $p'$ time-series are shown for RTI and KHRTI in Figs. \ref{fig6}(b) and \ref{fig6}(d), showing a higher Fourier amplitude in the case of KHRTI compared to RTI. The spectra for KHRTI are chaotic, revealing the presence of multiple time scales in the flow field due to competing KH and RT mechanisms. The spectra for RTI only show a peak near the zero frequency line, revealing that the time-period considered is over a nascent stage of the instability, i.e., without formation of typical coherent structures for RTI viz. spikes and bubbles. Next, we anticipate to derive more information about the spatio-temporal dynamics of the RTI and KHRTI by performing a proper orthogonal decomposition over the stored snapshots which are highly resolved in space and time.

\subsection{Proper orthogonal decomposition of RTI and KHRTI during onset}

\noindent In this section, we perform proper orthogonal decomposition (POD) to project spatio-temporal dynamics of the RTI and KHRTI on to a reduced number of deterministic eigenfunctions to completely (almost) represent the total energy \cite{berkooz1992coherent}, using the method of snapshots \cite{sirovich1987turbulence}. Traditionally, the velocity or vorticity field is used for the POD analysis \cite{sengupta2012instabilities} , here we use instead the space-time dependent disturbance pressure field $p'$ as it forms the basis of much of our earlier discussion on acoustic triggering of RTI and KHRTI. To perform POD for RTI and KHRTI computed with 4.2 billion points and 0.48 billion points, we use 1527 and 2166 snapshots of the ($x$, $y$)-plane respectively.  The full decomposition of $p'$ reads  

\begin{equation}
 p' (\vec{X}, t) = \sum_{m=1}^M a_m(t) \phi_m(\vec{X})
\end{equation}

\noindent where $M$ represents the total number of snapshots. The spatial modes $\{\phi_m\}$ can be obtained by solving an eigen problem for the correlation function $R_{ij} = \frac{1}{M} \int \int p'(\vec{X}, t_i) p'(\vec{X}, t_j) d^2 \vec{X}$, with $i, j = 1,2, ..., M$ defined as collocation points in the computational domain.  For this reason, they are often called eigenmodes. Having obtained the spatial modes in the chosen time range $[0,t_{max}]$, one can also calculate the POD amplitude function $\{a_m\}$ corresponding to the $m^{th}$ eigenmode using the orthogonal property of the POD basis as follows $a_i = \frac{\int \int_S p' \phi_i dS}{\int \int_S \phi^2_i dS}$. The first five modes account for a significant portion of the total contribution in both cases. Specifically, the RTI collectively contribute 95.30 percent of the total, indicating that most energy or variance is concentrated in these initial modes. Similarly, for the KHRTI, the first five modes contribute an even higher percentage of 98.61 percent, suggesting an even greater dominance of these modes.

\begin{figure*}[!ht]
\centering
\includegraphics[width=.88\textwidth]{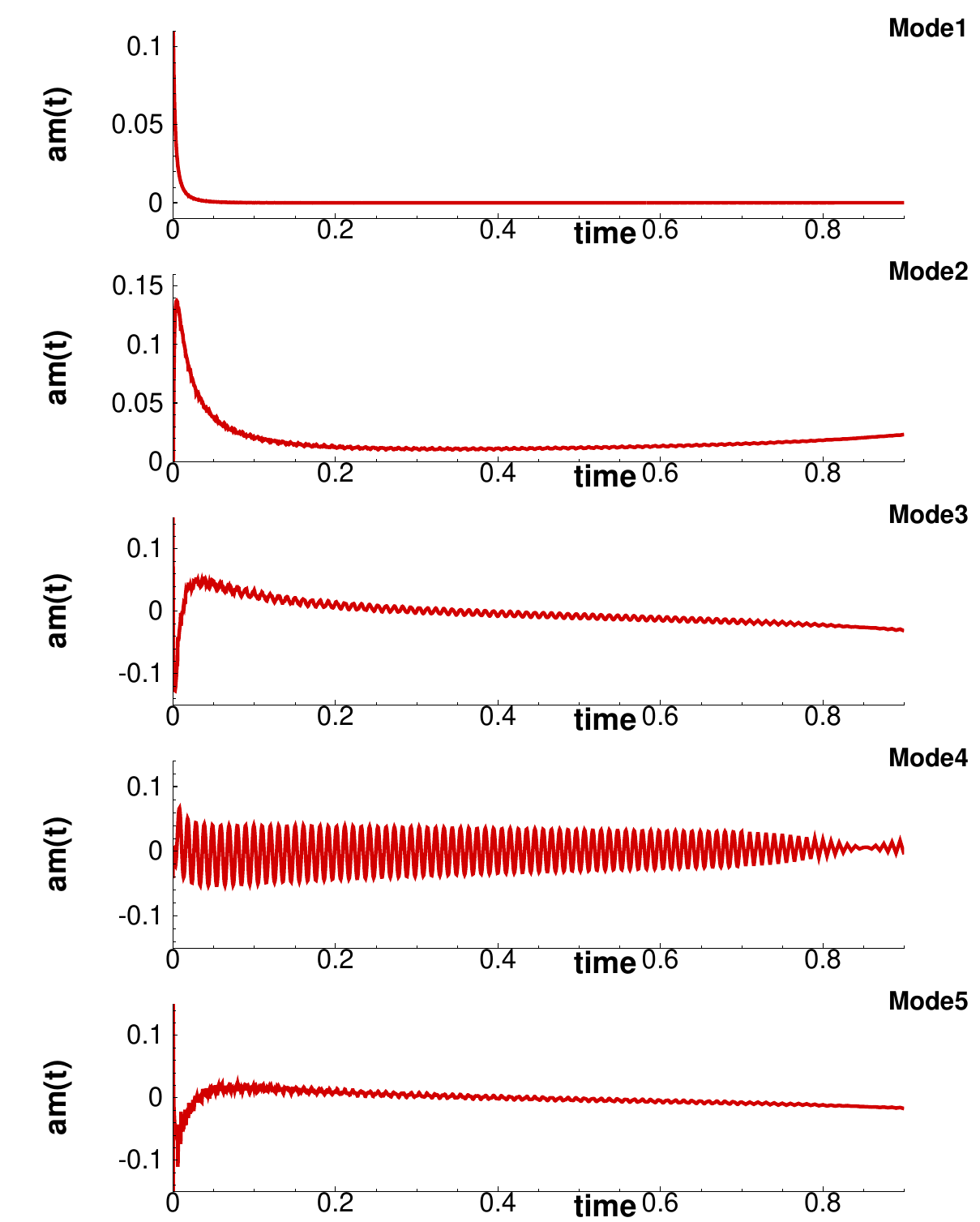}
\caption{Time-dependent amplitude functions of first five POD modes of RTI, evaluated over the time interval shown in Fig. \ref{fig6}.}
\label{fig7}
\end{figure*}

In Fig. \ref{fig7}, the first 5 time-dependent amplitude functions $\{a_i\}_{i=1}^5$ are shown for the computed RTI, over the time range $[0,0.9]$ which is the same as in Fig. \ref{fig6}. Modes 1 and 2 form a pair, which is termed as a regular mode \cite{sengupta2012instabilities} having an identical order of magnitude and time variation of the amplitude functions. Similarly, modes 3 and 5 form a regular mode pair. For flow past a cylinder, these regular modes have been found to satisfy the Stuart-Landau equation also \cite{noack2003hierarchy}. We cannot comment on the applicability of Stuart Landau equation for the RTI POD modes, but this poses as a topic for a potential future study. For eigenmodes $\phi_1$, $\phi_2$, $\phi_3$ and $\phi_5$, the time variation of the respective amplitude function resembles the time variation of $p'$ in Fig. \ref{fig10}. However, $a_4$ does not follow the time variation of the flow variable, $p'$, which is termed as anomalous mode in the literature \cite{lestandi2018multiple} for internal flows. These POD modes show a non-periodic shift with time, thus these are often also called \lq shift modes' \cite{noack2003hierarchy}. For the RTI computations, mode 4 acts in isolation, and is thus the anomalous or shift mode, which is responsible for the transition to turbulence of the instability \cite{sengupta2012instabilities}. 

\begin{figure*}[!ht]
\centering
\includegraphics[width=.88\textwidth]{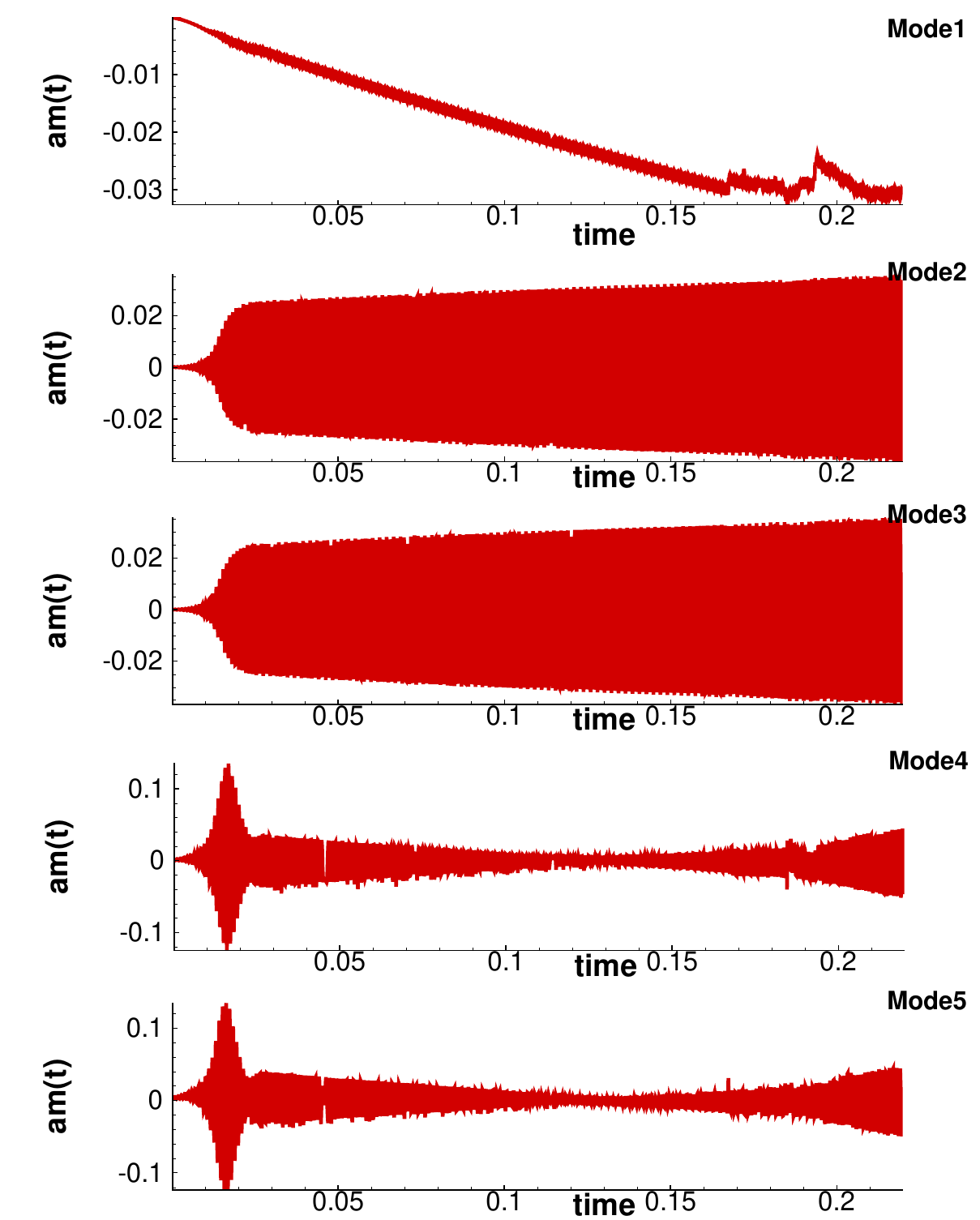}
\caption{Time-dependent amplitude functions of first five POD modes of KHRTI, evaluated over the time interval shown in Fig. \ref{fig6}.}
\label{fig8}
\end{figure*}

In Fig. \ref{fig8}, first five amplitude functions are shown for KHRTI, over the time interval of Fig. \ref{fig6}, i.e. from $[0, 0.24]$. Here too, just like in the case of RTI, we observe two pairs of regular modes, i.e. $a_2$, $a_3$ and $a_4$, $a_5$. However, only the regular mode pair $a_2$ and $a_3$ display time variation akin to that of $p'$ in Fig. \ref{fig6} for KHRTI. The other regular mode pair ($a_4$ and $a_5$) does not mimic the time variation of $p'$ for KHRTI, and whether either of these pairs satisfy the Stuart-Landau equation is to be determined in a future study. Correlating the POD modes with the instability modes will be a useful exercise for hydrodynamic instabilities, particularly with such highly resolved DNS data. For the KHRTI, mode 1 is identified as the anomalous or regular mode. This is considered as a mean field correction, i.e. showing the difference between time-averaged solution of unsteady NSE and solution of steady NSE \cite{noack2003hierarchy}. Interestingly in this reference, the dynamics of cylinder wake was influenced by the presence of this shift mode. For the KHRTI, instead of the vortex shedding in the cylinder flow, the anomalous mode may be associated with the travelling KH eddy. The role of anomalous modes in the growth of the RTI and KHRTI will be explored next by plotting its distribution in the ($x$, $y$)-plane. 

\begin{figure*}[!ht]
\centering
\includegraphics[width=.9\textwidth]{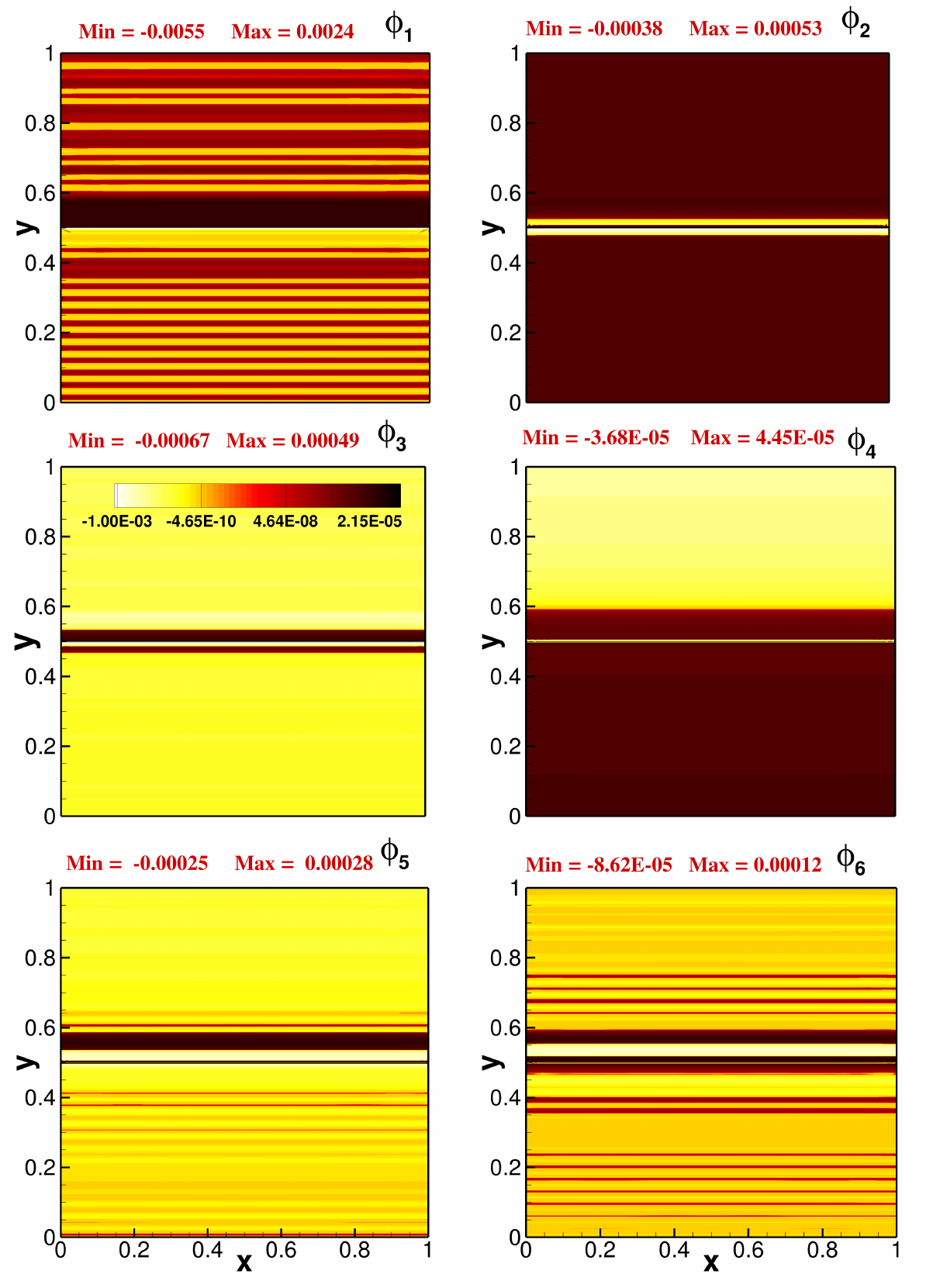}
\caption{Contour plots of first six spatial/eigenfunction modes ($\phi_1$ - $\phi_6$) from POD of RTI, evaluated over time interval shown in Fig. \ref{fig6}.}
\label{fig9}
\end{figure*}

\begin{figure*}[!ht]
\centering
\includegraphics[width=.85\textwidth]{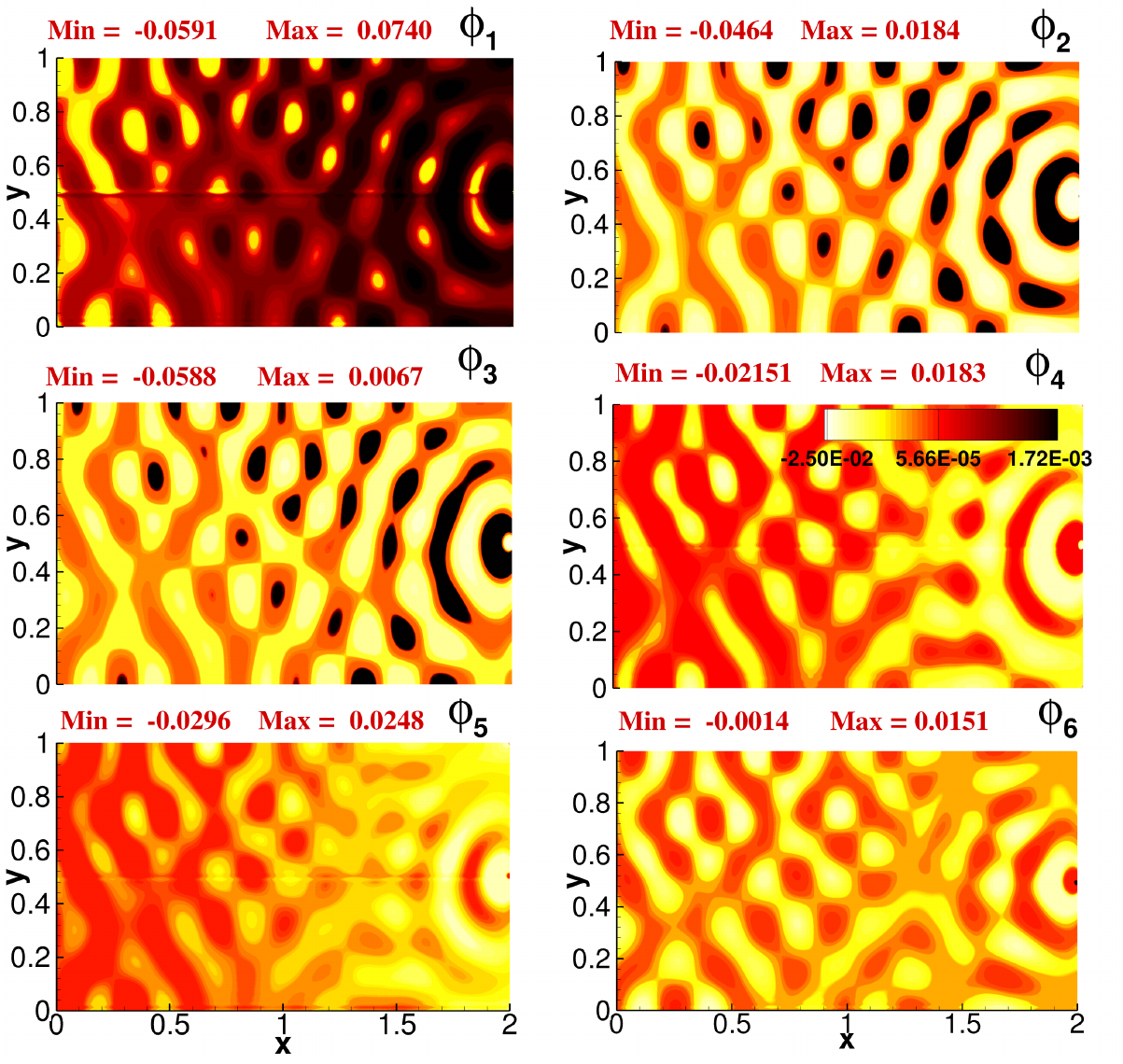}
\caption{Contour plots of first six spatial/eigenfunction modes ($\phi_1$ - $\phi_6$) from POD of KHRTI, evaluated over time interval shown in Fig. \ref{fig6}.}
\label{fig10}
\end{figure*}

In Fig. \ref{fig9}, the leading six spatial modes of the data set for RTI computed over the time interval of Fig. \ref{fig6} are shown in the ($x$, $y$)-plane. The first two modes form a regular pair and the contour plots are complementary to each other. The first mode captures the travelling low wavenumber pressure pulses from the interface to either wall while the second mode focuses on the interfacial pressure perturbation field. A similar observation can be made about the regular mode pair, $\phi_3$ and $\phi_5$, wherein the third mode captures the interface pressure perturbation while the fifth mode shows the signature of the convecting pressure pulses. Mode-4, on the other hand, which displayed the anomalous behavior in Fig. \ref{fig7}, is in isolation. It does not form a pair with any other mode. To establish this, mode-6 has been displayed which again captures convecting pressure pulses forming a regular mode pair with mode 7 (not shown). In fact, $\phi_4$ is representative of the pressure field due to the imposed density gradient. This is, thus, an attribute of variation of the flow in a slow time scale, i.e., the mode is a slowly varying wave travelling downstream. The role of the shift mode in RTI is similar to the one observed in the context of vortex shedding in flow past a cylinder or appearance of limit cycle oscillation in a lid-driven cavity \cite{sengupta2012instabilities}. 

Figure \ref{fig10} displays the leading six spatial modes in the ($x$, $y$)-plane for the KHRTI, computed over the time interval of Fig. \ref{fig6}. Here, the paired up regular modes, $\phi_2$ and $\phi_3$ are complementary as can be seen from the contour plots. Similarly, the two modes, $\phi_4$ and $\phi_5$ are mutually orthogonal. Mode-1, on the other hand, is the anomalous mode (found in isolation), which shows the closest resemblance to the radial pressure pulse propagation of Fig. \ref{fig3}. Here too, the shift mode represents the transient effects. Since the time interval used for the POD captures the early onset of the KHRTI, rapid transients are observed in the flow. We expect the transience to be captured by the anomalous modes, and in fact, here, mode-1 which is the most dominant eigenmode is responsible for representing the unsteady dynamics of the KHRTI. 

\subsection{Rotationality and budgets for translational kinetic energy and compressible enstrophy: RTI versus KHRTI}

\noindent The discussion and analysis so far has been restricted to the spatio-temporal evolution of disturbance pressure pulses. However, the origin of RTI and other hydrodynamic instabilities is often conjectured to be due to generation of baroclinic vorticity, $-\frac{1}{\rho} (\nabla p \times \nabla \rho)$ \cite{chandrasekhar1961hydrodynamic}. For shear-driven KHRTI, on the other hand, the onset process involves creation and growth of Kelvin-Helmholtz eddies. Thus, rotationality (or vorticity, $\vec{\omega}$) has a central role in the development of RTI and KHRTI, and in this section, we will explore the vorticity dynamics of the disturbance field. We will also show the energy budget involved in RTI and KHRTI by comparing disturbance pressure energy, turbulent kinetic energy, and disturbance enstrophy contributions. Finally, the central mechanism during onset will be brought out by the application of the compressible enstrophy transport equation (CETE). 

\begin{figure*}[!ht]
\centering
\includegraphics[width=.9\textwidth]{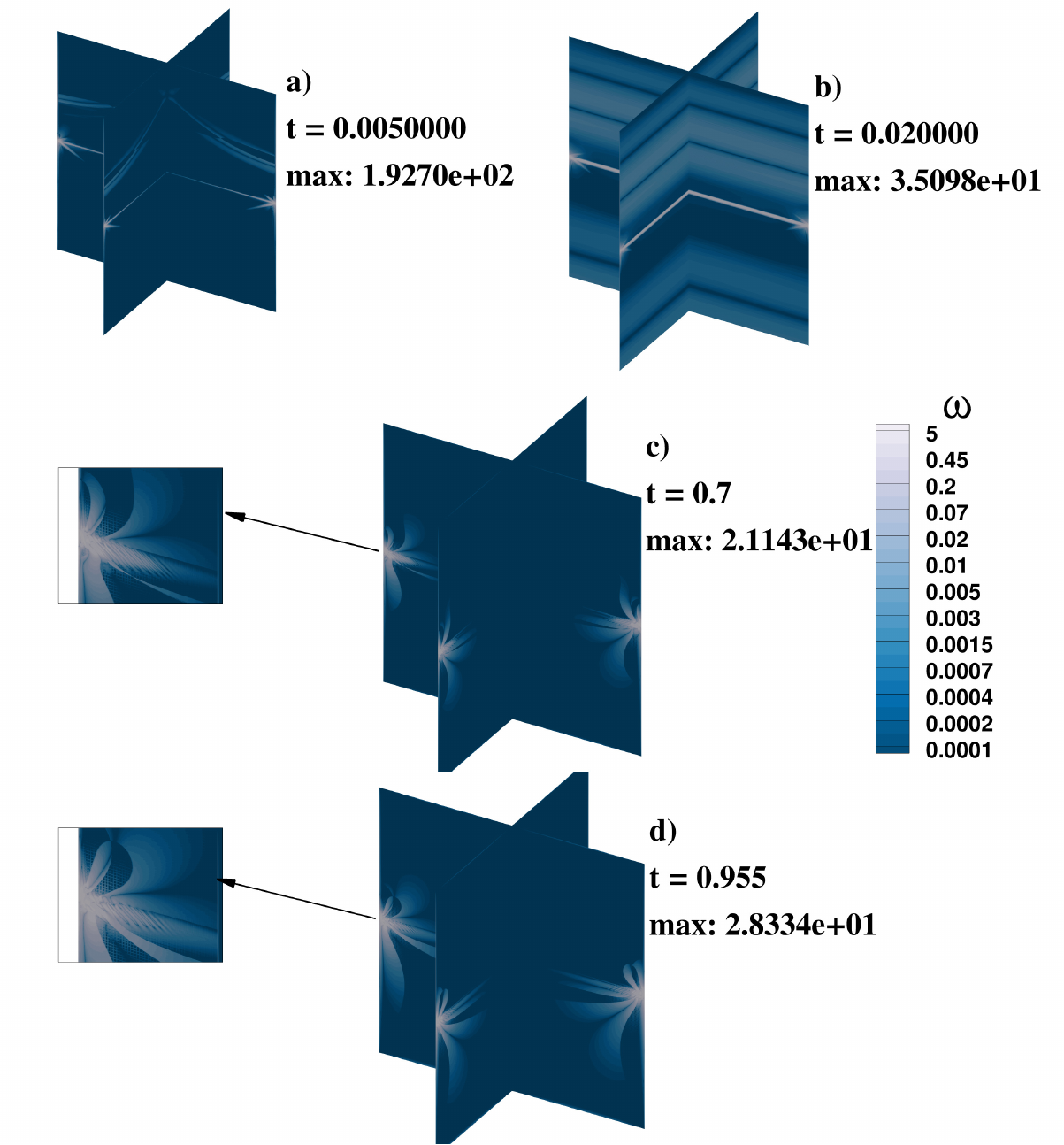}
\caption{Disturbance vorticity contours shown across mid-z and mid-x planes for RTI at indicated time instants. }
\label{fig11}
\end{figure*}

In Fig. \ref{fig11}, the disturbance vorticity, $\omega'_d$ is shown for the indicated time instants along the ($x$, $y$)- and ($y$, $z$)-planes to demonstrate the onset of RTI. The early onset at $t = 0.005$, captured in Fig. \ref{fig11}(a), shows the development of baroclinic vorticity at the junctures of the side-walls with the original interface. The misalignment of the pressure and density gradients is found to be maximum at the juntion points. There is a non-negligible contribution arising due to the travelling compression and rarefaction fronts in the interface normal direction, as seen in Figs. \ref{fig11}(a) and \ref{fig11}(b). By a later time of $t = 0.7$ in Fig. \ref{fig11}(c), the disturbance vorticity collates at the junction points with clear evidence of the baroclinic torque (as viewed in the zoomed inset plot). The onset of RT billows is most clearly evident in the latest time instant in Fig. \ref{fig11}(d). There is a qualitative match between flow structures noted for $p'$ in Fig. \ref{fig2} and $\omega'_d$. The origin of the vorticity can be traced back to the travelling acoustic pulses from the interface, which trigger the $\nabla p$ responsible for baroclinic torque generation. 

\begin{figure*}[!ht]
\centering
\includegraphics[width=.9\textwidth]{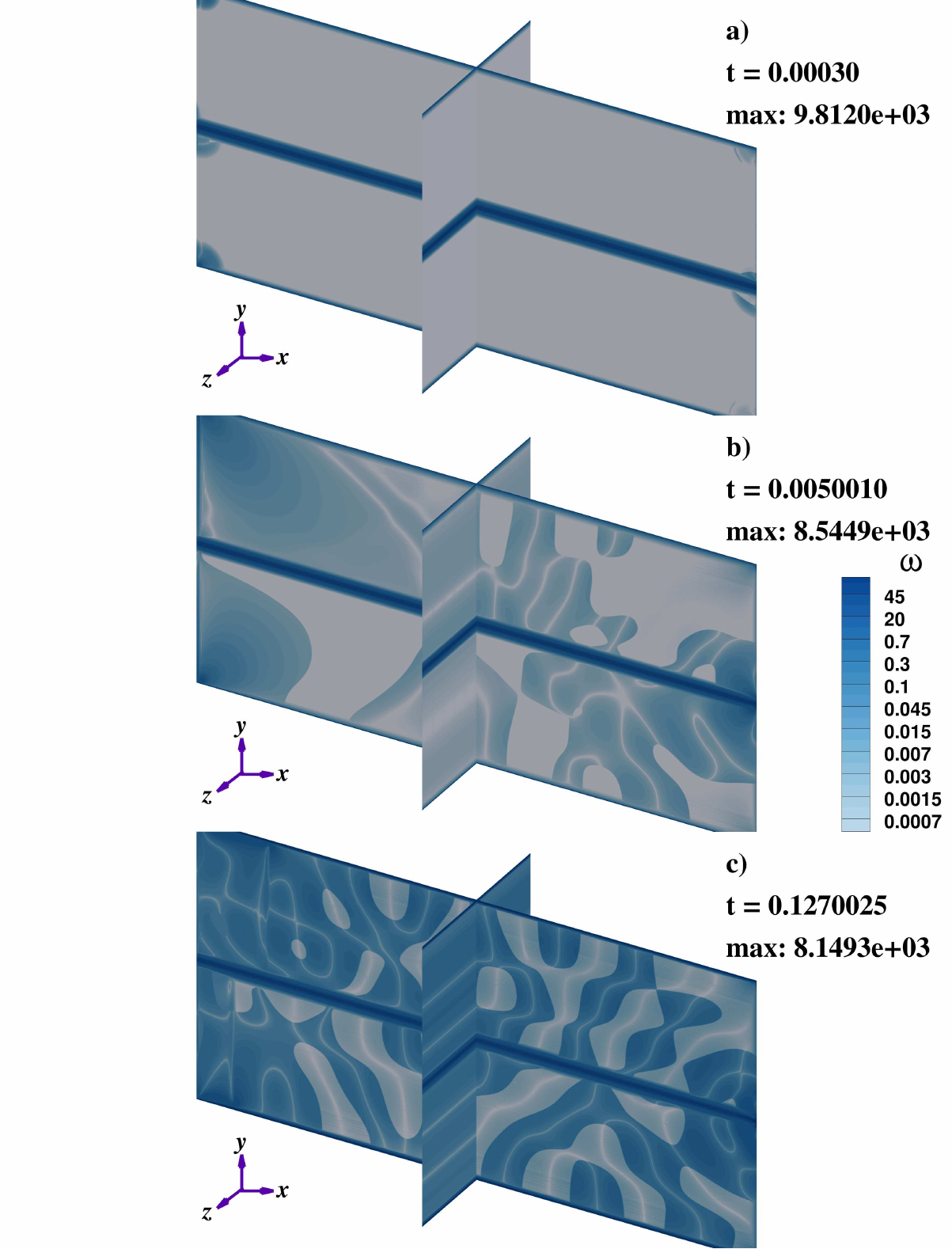}
\caption{Disturbance vorticity contours shown across mid-z and mid-x planes for KHRTI at indicated time instants. }
\label{fig12}
\end{figure*}

In Fig. \ref{fig12}, the disturbance vorticity, $\omega'_d$ is shown for the indicated time instants along the ($x$, $y$)- and ($y$, $z$)-planes to demonstrate the onset of KHRTI. Here also, one notes a qualitative match between contours of $\omega'_d$ and $p'$ in Fig. \ref{fig3}. However, the radial pressure pulses travelling in the upstream direction are much more coherent than the corresponding disturbance vorticity. This reaffirms that the trigger for the instability is an acoustic one. It is the travelling pressure pulses in radial and interface normal directions which leads to the misalignment of $\nabla p$ and $\nabla \rho$, and hence leads to formation of baroclinic vorticity. This has been shown here for the first time, and challenges the notion that the genesis of RTI/KHRTI is due to baroclinic torque - rather the acoustics trigger the vorticity generation.

\begin{figure*}[!ht]
\centering
\includegraphics[width=.92\textwidth]{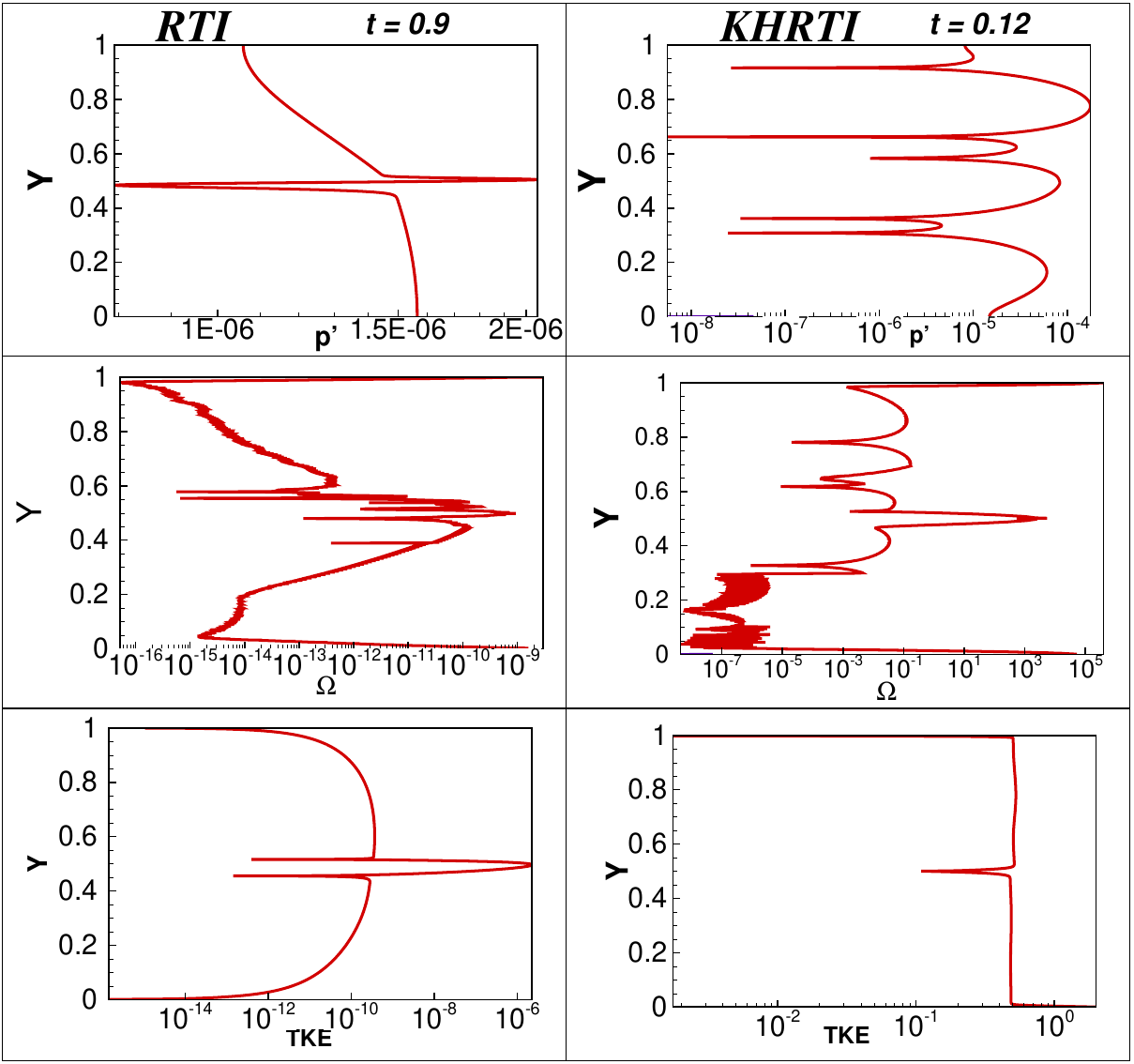}
\caption{Wall-normal distribution of $p'$, enstrophy and the turbulent kinetic energy are plotted for mid-x and mid-z planes at indicated time instant for the RTI and KHRTI}
\label{fig13}
\end{figure*}

Similar to turbulent kinetic energy budget \cite{sengupta2020effectsa}, which tracks transport and redistribution of translational energy, enstrophy provides an effective measure of rotational energy in a 3D viscous flow. It is defined as the self-dot product of the vorticity vector, $\Omega = \vec{\omega}\cdot \vec{\omega}$. We will compare the wall-normal energy contributions from disturbance pressure, translational energy (turbulent kinetic energy), and rotational energy (disturbance enstrophy) in Fig. \ref{fig13} along the mid-stream and mid-span plane. In the left frames, the quantities are plotted for RTI, and in right frames, we have the KHRTI values. Overall, for KHRTI, maximum contribution is from rotational energy, i.e., $\Omega$, whereas for RTI, $p'$ has the largest contribution. The peak location for all quantities is at the interface, except for $p'$ for KHRTI, wherein the entire wall-normal extent shows multiple peaks. This can be explained by the radially propagating pressure pulses in the ($x$, $y$)-plane for KHRTI in Fig. \ref{fig3}. The translational turbulent kinetic energy is understandably orders of magnitude higher for KHRTI compared to RTI due to the imposed shear. The acoustic trigger is of similar order of magnitude for RTI and KHRTI, but it has larger contribution in generation of baroclinic torque in RTI compared to KHRTI, wherein the added KH mechanism (showing up in the turbulent kinetic energy) has a role too. 

The transport equation of compressible enstrophy \cite{suman2022novel} serves as a valuable tool for analyzing the generation, distribution, and evolution of enstrophy during the transition to turbulence in various internal \cite{sengupta2020effectsa} and external flows \cite{sengupta2024separation}. In the context of RTI, studies have demonstrated that energy is redistributed from small-scale billowing structures near the side walls to large-scale vortical formations at the interface's center \cite{zhou2024hydrodynamic}. Additionally, KH eddies, characterized by localized vorticity whirls, facilitate energy transfer from large-scale eddies to the small-scale plumes typical of RTI dynamics \cite{sengupta2023effects}. This interplay highlights the critical role of vorticity and enstrophy in understanding the energy cascade and flow evolution in complex instability-driven systems.

The formulation of the CETE from the compressible NSE was introduced in an earlier work \cite{suman2022novel}. Applications of CETE to RTI and coarse-resolution KHRTI problems \cite{sengupta2023role, sengupta2022thermally, sengupta2023effects} have highlighted the dominant role of viscous terms in enstrophy evolution for buoyancy-dominated flows. This finding challenges the traditional emphasis on baroclinic terms as the primary trigger for the RTI \cite{chandrasekhar1961hydrodynamic, lawrie2010rayleigh}. For advection-dominated flows, however, the vortex stretching term emerges as a significant contributor to enstrophy evolution alongside viscous effects \cite{sengupta2022thermally, sengupta2023effects}. Despite this, enstrophy growth often traces back to baroclinic contributions \cite{sengupta2023multi}, especially once coherent structures (spikes, bubbles, KH eddies) span the interface. Investigations into inviscid vorticity generation mechanisms have underscored the dominance of baroclinic effects during the early stages of Richtmyer-Meshkov instabilities \cite{pereira2020effect}. In contrast, analyses of enstrophy budgets for the same instability \cite{zhou2020dependence} reveal that, in fully 3D flows, baroclinic contributions are negligible, while the vortex stretching term is two orders of magnitude more significant. This section applies CETE to highly resolved DNS of RTI and KHRTI. The objective is to investigate the enstrophy dynamics governing the initial growth phase of these instabilities, providing insights into the mechanisms responsible during onset. The constituent terms of the CETE are as follows \cite{suman2022novel}:

	\begin{equation}
		\begin{aligned}
			\frac{D{\Omega }}{Dt}
			= & \; 2\vec{\omega } \cdot \left[(\vec{\omega} \cdot {\nabla}) \vec{V}\right] - 2({\nabla} \cdot \vec{V}) \Omega \\
			& + \left(\frac{2}{\rho^{2}}\right) \vec{\omega } \cdot \left[\left({\nabla \rho} \times {{\nabla p}}\right)\right] -\left(\frac{2}{\rho^{2}}\right)\vec{\omega } \cdot \left[{\nabla \rho} \times {\nabla} \left(\lambda ({\nabla} \cdot \vec{V})\right)\right] \\
			& +\left(\frac{4}{\rho}\right)\vec{\omega } \cdot \left[{\nabla} \times \left[{\nabla} \cdot \left(\mu S \right)\right]\right] -\left(\frac{4}{\rho^2}\right)\vec{\omega } \cdot \left({\nabla \rho} \times ({\nabla} \cdot \left(\mu S\right)) \right)
		\end{aligned}
		\label{CETE}
	\end{equation}
	
The various terms of Eq. \eqref{CETE} are as follows:

\begin{itemize}
		
\item $2\vec{\omega } \cdot \left[(\vec{\omega} \cdot {\nabla}) \vec{V}\right]$ : Contribution to enstrophy due to vortex stretching (T1). 
		
\item $({\nabla} \cdot \vec{V}) \Omega$: Enstrophy growth/decay due to compressibility (T2). 
		
\item $\left(\frac{1}{\rho^{2}}\right) \vec{\omega } \cdot \left[\left({\nabla \rho} \times {{\nabla p}}\right)\right]$: Contribution from baroclinic term due to misalignment of gradients of pressure and density (T3). 
		
\item $\left(\frac{1}{\rho^{2}}\right)\vec{\omega } \cdot \left[{\nabla \rho} \times {\nabla} \left(\lambda ({\nabla} \cdot \vec{V})\right)\right]$ : Contribution due to misalignment of vorticity and bulk viscosity gradients (T4).
		
\item $\left(\frac{1}{\rho}\right)\vec{\omega } \cdot \left[{\nabla} \times \left[{\nabla} \cdot \left(\mu S \right)\right]\right]$: Diffusion of enstrophy due to viscous action (T5).
		
\item $\left(\frac{1}{\rho^2}\right)\vec{\omega } \cdot \left({\nabla \rho} \times ({\nabla} \cdot \left(\mu S\right)) \right)$: Contribution due to misalignment of gradients of density and divergence of viscous stresses (T6).
\end{itemize}

\begin{figure*}[!ht]
\centering
\includegraphics[width=.95\textwidth]{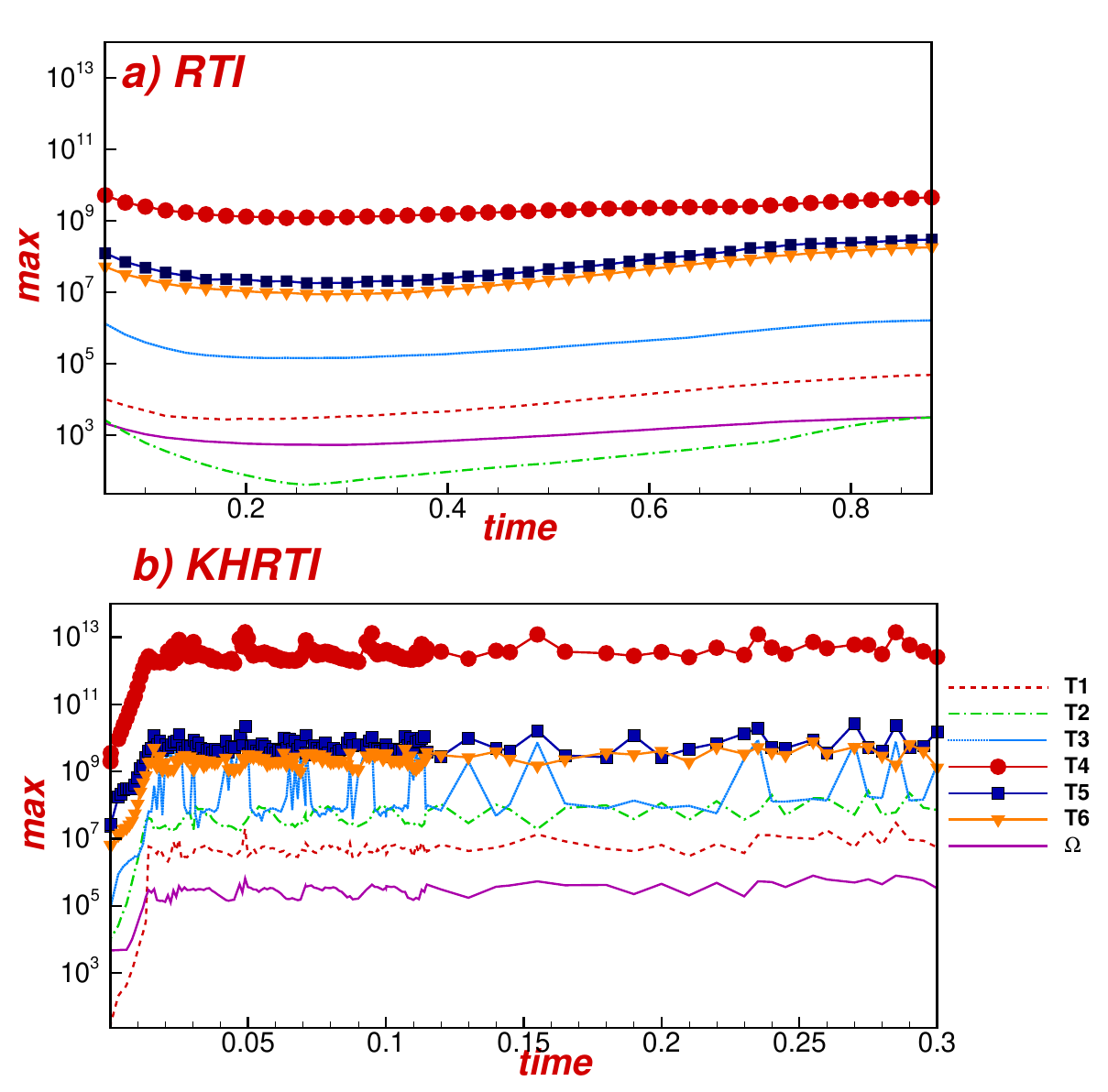}
\caption{Evolution of maximum CETE budget terms (T1-T6) and enstrophy shown for a) RTI and b) KHRTI. The $y$-scale is logarithmic.}
\label{fig14}
\end{figure*}

In Fig. \ref{fig14}, evolution of the maximum CETE budget terms is compared for RTI and KHRTI. In Fig. \ref{fig14}(a) the CETE budget evolution for the RTI is displayed, while Fig. \ref{fig14}(b) shows the evolution for the KHRTI. For both RTI and KHRTI, viscous terms (T4, T5, and T6) play a dominant role in the enstrophy growth rate, right from the onset. T4 arises from the bulk viscosity, and due to the imposed non-zero bulk viscosity \cite{ash1991second} in the present formulation, understandably it has the largest contribution to CETE. The baroclinic term, T3 has the next largest contribution for both RTI and KHRTI. As the computed DNS only captures the onset stages, without the formation of characteristic RT plumes and/or KH eddies, it is expected that at later stages T3 will have a larger contribution. For similar reasons, the vortex stretching term T1 for KHRTI is sub-dominant compared to viscous stress terms. The overall magnitude of budget terms is four orders of magnitude higher for KHRTI than its RTI counterpart. The higher disturbance levels in KHRTI is well-established from the discussion so far on the disturbance pressure field. Interestingly, for KHRTI, term T2 which is enstrophy growth due to compressibility effects is significant compared to the CETE budget of RTI. This can be attributed to the existence of pressure pulses or wavefronts that move in upstream and interface-normal directions, eventually filling the entire computational domain. These contribute to more intense compression and rarefaction, influencing the compressibility term. For RTI, on the other hand, we start with a quiescent condition having Mach number of 0.003, and early onset does not contribute significantly to the dilatation term. At onset, vortex stretching term T1 initially shows the least contribution for KHRTI. However, once the KH mechanism sets in, a sudden spike in the vortex stretching is observed, and it is expected to dominate baroclinicity and viscous contributions once KH eddies span the interface. 

\section{Summary and conclusions}
\label{sec4}

\noindent The present study compares the onset process of the Rayleigh-Taylor instability (RTI) and Kelvin-Helmholtz Rayleigh-Taylor instability (KHRTI) using highly-resolved direct numerical simulations. These simulations involve two computational domains, illustrated in Fig. \ref{fig1}, where air is considered at two different temperatures (densities) and two velocities (in case of KHRTI) in the two halves of the computational box. The simulations are performed using 4.19 billion and 0.48 billion mesh points respectively, for the RTI and KHRTI. The compressible Navier-Stokes equations are solved with highly accurate, dispersion relation preserving compact schemes \cite{sengupta2017hybrid, sagaut2023global} for discretization and the parallel programming is performed using an error-free non-overlapping parallel algorithm \cite{sundaram2023non}. The accuracy of the numerical framework allows for the capture of pressure pulses in infrasonic and ultrasonic range. 

The onset of RTI and KHRTI is compared by tracking the disturbance pressure contours in Figs. \ref{fig2} and \ref{fig3}. For RTI, pressure pulses are found to travel in the interface-normal direction. For the KHRTI, the pressure pulses travel in the upstream direction \cite{joshi2024highly} radially in the ($x$, $y$)-plane. In the ($y$, $z$)-plane, however, the pressure pulses travel in the interface-normal direction, as in RTI. Thus, the imposed advection in the $x$-direction in the KHRTI interacts with the convecting pressure pulses, resulting in a radial propagation of the pressure pulses. The corresponding spectra of the disturbance pressure in Fig. \ref{fig4} reveal for the RTI distinct modal, nonmodal and high wavenumber signatures. On the other hand, the KHRTI spectrum shows distinct modal and nonmodal components only when convection-dominated RT mechanism is prominent (immediately after onset). There is no distinct high wavenumber component in the spectrum, $p'$ is distributed over a wide range of wavenumbers. Anomalous behavior in the spectrum of KHRTI is observed whenever there is an interaction between the radial pressure pulse travelling upstream in the ($x$, $y$)-plane and interface-normal pressure pulse in ($y$, $z$)-plane, as described in Fig. \ref{fig5}. The time evolution of $p'$ in Fig. \ref{fig6} reveals that it has a higher magnitude at the sides of the computational domain compared to the centre for both RTI and KHRTI, which is expected as we have only computed the onset stage before coherent structure formation along the interface. 

The proper orthogonal decomposition (POD) of RTI and KHRTI in Figs. \ref{fig7}-\ref{fig10} (based on disturbance pressure) shows the presence of two pairs of regular modes. The time variation of amplitude function of these eigenmodes is similar to the time variation of $p'$. An anomalous (or shift) mode \cite{noack2003hierarchy} is also observed for both RTI and KHRTI which explains the transient behavior of the flow, involving an aperiodic time variation which is completely different from the flow variable under consideration. The regular modes are mutually complementary, with one describing interfacial phenomena and the other describing convecting/radial pressure pulse propagation in the rest of the domain. The anomalous mode, on the other hand, is representative of temporal dynamics of the flow field. 

The vorticity dynamics of the RTI and KHRTI are explored in Figs. \ref{fig11} and \ref{fig12}, establishing that the baroclinic vorticity generation is a consequence of the acoustic trigger. The travelling rarefaction and compression fronts amplify the misalignment between pressure and density gradients, thereby leading to the origin of baroclinic torque. Thus, it is the acoustics which trigger the instability, not the baroclinic vorticity (which is a consequence of the acoustic waves). This has been demonstrated here unequivocally here for the first time. The wall-normal energy budget is examined along the mid-stream and mid-span plan in Fig. \ref{fig13}, revealing that enstrophy (or rotational energy) is dominant compared to the translational kinetic energy or pressure energy for KHRTI. However, for early onset of RTI, pressure energy is the highest contributor. For KHRTI, with an imposed streamwise shear, the kinetic energy is several orders of magnitude higher than RTI which starts from quiescent ambience. Thus, pressure perturbations are all the more important in triggering RTI compared to the KHRTI, although both are triggered by the acoustics.

Finally, application of the compressible enstrophy transport equation in Fig. \ref{fig14} reveals that viscous bulk viscosity term is predominant due to imposed non-zero bulk viscosity, closely followed by other viscous contributions. The baroclinic term is sub-dominant for this onset stage for both RTI and KHRTI. The enstrophy growth due to compressibility is much higher in case of KHRTI compared to RTI, which can be explained by the compression and rarefaction pulses propagating over the entire span of the computational domain, compared to the convecting interface-normal pulses in case of RTI. 

The present work serves as a highly resolved database for RTI and KHRTI calculations which can be used to conduct pressure perturbation based analysis and track vorticity dynamics during early onset, across various applications. We intend to extend our analysis for the later stages of RTI and KHRTI and to explore whether the POD modes obtained have a correlation with the instability modes obtained from the application of the Stuart-Landau equation \cite{sengupta2012instabilities}.

\section*{Credit authorship contribution statement}
Bhavna Joshi: Data curation, Investigation, Methodology, Visualization, Writing - review \& editing. Aditi Sengupta: Conceptualization and Supervision, Formal analysis, Funding acquisition, Investigation, Methodology, Project administration, Writing- Original draft preparation. Lucas Lestandi: Supervision, Funding acquisition, Investigation, Project administration, Writing- review \& editing. Yassin Ajanif: Investigation, Writing - review \& editing. 

\section*{Declaration of competing interest}
The authors declare that they have no known competing financial interests or personal relationships that could have appeared to influence the work reported in this paper.

\section*{Data Availability}
The data that support the findings of this study are available from Aditi Sengupta (aditi@iitism.ac.in) upon reasonable request. 


\bibliographystyle{elsarticle-num} 
\bibliography{references}




\end{document}